\documentclass[12pt,fleqn]{article}
\usepackage{amsfonts,ulem,url}
\usepackage{amssymb}
\usepackage{amsmath}
\usepackage{amstext,verbatim,graphicx,ifthen,multirow,amsthm}
\usepackage{bm}

\renewcommand{\mathbf}{\boldsymbol}
\renewcommand{\thepage}{}
\renewcommand{\appendix}{\footnotesize\parindent 0cm\setcounter{equation}{0}
\renewcommand{\theequation}{A.\arabic{equation}}
\setcounter{lemma}{0}\renewcommand{\thelemma}{A.\arabic{lemma}}}
\newcommand{\rif}{{\rm if}\ }
\newcommand{\itt}{{\rm itt}}
\newcommand{\neyman}{{\rm neyman}\ }
\newcommand{\robust}{{\rm robust}\ }
\newcommand{\nfe}{N_{{\rm f}\ttt}}
\newcommand{\nme}{N_{{\rm m}\ttt}}

\newcommand{\nfn}{N_{{\rm f}\ccc}}
\newcommand{\nmn}{N_{{\rm m}\ccc}}
\newcommand{\fen}{\frac{1}{N}}
\newcommand{\ow}{\overline{W}}

\newcommand{\tc}{{\rm c}}

\newcommand{\ta}{{\rm a}}
\newcommand{\tn}{{\rm n}}

\newcommand{\oyobse}{\overline{Y}^{\mathrm{obs}}_{\ttt}}
\newcommand{\oyobsn}{\overline{Y}^{\mathrm{obs}}_{\tc}}

\newcommand{\aand}{\hskip1cm {\rm and}\ \ }
\newcommand{\yoi}{Y_{i}^{\obs}}

\newcommand{\sscs}{S^2_{\tc}}
\newcommand{\ssts}{S^2_{\ttt}}
\newcommand{\iwc}{{i:W_i=\cc}}
\newcommand{\iwt}{{i:W_i=\ct}}

\newcommand{\fff}{{\rm f}}
\newcommand{\mmm}{{\rm m}}

\newcommand{\scs}{s^2_{\tc}}
\newcommand{\sts}{s^2_{\ttt}}

\newcommand{\mms}{\mathbb{S}}
\newcommand{\mmx}{\mathbb{X}}

\newcommand{\nf}{N_{\rm f}}
\newcommand{\nm}{N_{\rm m}}
\newcommand{\mmy}{\mathbb{Y}}
\newcommand{\hmmv}{\hat{\mathbb{V}}}
\newcommand{\sumn}{\sum_{i=1}^N}
\newcommand{\cc}{0}
\newcommand{\ct}{1}

\newcommand{\yin}{Y_i(0)}
\newcommand{\yie}{Y_i(1)}

\newcommand{\ccc}{{\rm c}}
\newcommand{\ttt}{{\rm t}}
\newcommand{\nt}{N_\ttt}
\newcommand{\nc}{N_\ccc}

\newcommand{\mmc}{\mathbb{C}}
\newcommand{\pop}{{\rm pop}}

\newcommand{\indep}{\perp\!\!\!\perp}

\def\monthname{\ifcase\month\or
  January\or February\or March\or April\or May\or June\or July\or
  August\or September\or October\or November\or December\fi}
\renewcommand{\thepage}{[\arabic{page}]}
\numberwithin{equation}{section}

\newcommand{\been}{{\bf 1}}

\newcommand{\ggc}{G} 

\newcommand{\pr}{{\rm pr}}
\newcommand{\obs}{{\rm obs}}

\newcommand{\oy}{\overline{Y}}
\newcommand{\ox}{\overline{X}}

\newcommand{\bge}{\begin{equation}}
\newcommand{\ene}{\end{equation}}
\newcommand{\by}{{\mathbf{Y}}}
\newcommand{\bw}{{\mathbf{W}}}

\newcommand{\bx}{{\mathbf{X}}}

\newcommand{\ols}{{\rm ols}}
\newcommand{\ave}{{\rm ave}}

\newcommand{\mmv}{{\mathbb{V}}}

\newcommand{\mme}{{\mathbb{E}}}

\def\monthname{\ifcase\month\or
January\or February\or March\or April\or May\or June\or
July\or August\or September\or October\or November\or December\fi}
\renewcommand{\appendix}{\small\parindent 0cm\setcounter{equation}{0}
\renewcommand{\theequation}{A.\arabic{equation}}
\setcounter{lemma}{0}\renewcommand{\thelemma}{A.\arabic{lemma}}
\setcounter{theorem}{0}\renewcommand{\thetheorem}{A.\arabic{theorem}}}
\begin{document}

\title{The Econometrics of Randomized Experiments\thanks{{\small We are
grateful for comments by Esther Duflo.}} }
\author{ Susan Athey\thanks{{\small Professor of
Economics, Graduate School of Business, Stanford University, and NBER,
athey@stanford.edu. }} \and Guido W. Imbens\thanks{{\small Professor of
Economics,
Graduate School of Business, Stanford University, and NBER,
imbens@stanford.edu. }} }
\date{ Current version \ifcase\month\or
January\or February\or March\or April\or May\or June\or
July\or August\or September\or October\or November\or December\fi \ \number%
\year\ \  }
\maketitle


\begin{abstract}
In this chapter, we present econometric and statistical  methods for analyzing randomized experiments. For basic experiments we stress randomization-based inference as opposed to sampling-based inference. In randomization-based inference, uncertainty in estimates arises naturally from the random assignment of the treatments, rather than from hypothesized sampling from a large population.
We show how this perspective relates to regression analyses for randomized experiments.
We discuss the analyses of stratified, paired, and clustered randomized experiments, and we stress the general efficiency gains from stratification.
We also discuss complications in randomized experiments such as non-compliance. In the presence of non-compliance we contrast intention-to-treat analyses with instrumental variables analyses allowing for general treatment effect heterogeneity.
We consider in detail estimation and inference for heterogenous treatment effects in settings with (possibly many) covariates. These methods allow researchers to explore heterogeneity by identifying subpopulations with different treatment effects while maintaining the ability to construct valid confidence intervals. We also discuss optimal assignment to treatment based on covariates in such settings.
Finally, we discuss estimation and inference in experiments in settings with interactions between units, both in general network settings and in settings where the population is partitioned into groups with all interactions contained within these groups.
\end{abstract}

\begin{center}
\end{center}



\baselineskip=20pt\newpage \setcounter{page}{1}\renewcommand{\thepage}{[%
\arabic{page}]}\renewcommand{\theequation}{\arabic{section}.%
\arabic{equation}}

\section{Introduction}
\label{section:introduction}

Randomized experiments have a long tradition in agricultural and biomedical settings. 
In economics they have a much shorter history. Although there have been notable experiments over the years,  such as the RAND health care experiment (Manning, Newhouse,  Duan,  Keeler and 
Leibowitz, 1987, see the general discussion in Rothstein and  von Wachter, 2016) and the Negative Income Tax experiments (e.g., Robins, 1985),  it is only recently that there has been a 
large number of randomized experiments in economics, and development economics in particular. See Duflo,  Glennerster, and  Kremer (2006) for a survey. As digitization lowers the cost of conducting 
experiments, we may expect that their use may increase further in the near future.  In this chapter we discuss some of the statistical methods that are important 
for the analysis and design of randomized experiments. 

Although randomized experiments avoid many of the challenges of observational studies for causal inference, there remain
a number of statistical issues to address in the design and analysis of experiments.  Even in the simplest case with observably
homogenous, independent subjects, where the experiment is evaluated by comparing sample means for the treatment
and control group, there are questions of how to conduct inference about the treatment effect.  When there are observable
differences in characteristics among units, questions arise about how best to design the experiment and how to account for imbalances in 
characteristics between the treatment and control group in analysis.  In addition, it may be desirable to understand how
the results of an experiment would generalize to different settings.  One approach to this is to estimate heterogeneity in
treatment effects; another is to reweight units according to a target distribution of characteristics.  Finally, statistical issues
arise when units are not independent, as when they are connected in a network.  In this chapter,
we discuss a variety of methods for addressing these and other issues.

A major theme of the chapter is that we recommend using statistical methods that are 
directly justified by randomization, in contrast to the more traditional sampling-based approach that is
commonly used in econometrics.  In essence, the sampling based approach considers the treatment assignments
to be fixed, while the outomes are random.  Inference is based on the idea that the subjects are a random sample
from a much larger population.  In contrast, the randomization-based approach takes the subject's potential outcomes
(that is, the outcomes they would have had in each possible treatment regime) as fixed, and considers the assignment
of subjects to treatments as random.   Our focus on randomization follows the spirit of 
Freedman (2006, p. 691),  who  wrote: ``Experiments should be analyzed
as experiments, not as observational studies. A simple comparison of rates might be just
the right tool, with little value added by `sophisticated' models.'' Young (2016) has recently applied randomization-based
methods in development economics.

As an example of how the randomization-based approach matters in practice, 
we show that methods that might seem natural to economists in the conventional sampling paradigm
(such as controlling for observable heterogeneity using a regression model) require additional assumptions in order to
be justified.  Using the randomization-based approach suggests alternative methods, such as placing the data
into strata according to covariates, analyzing the within-group experiments, and averaging the results.  This is directly justified by randomization of the treatment assignment, and does not require any additional
assumptions.  

Our overall goal in this chapter is to collect in one place some of the most important statistical methods for analyzing and designing randomized experiments. We will start by discussing some general aspects of randomized experiments, and why they are widely viewed as providing the most credible evidence on causal effects. We will then present a brief introduction to causal inference based on the potential outcome perspective. Next we discuss the analysis of the most basic of randomized experiments, what we call completely randomized experiments where, out of a population of size $N$, a set of $\nt$ units are selected randomly to receive one treatment and the remaining $\nc=N-\nt$ are assigned to the control group. We discuss estimation of, and inference for, average as well as  quantile treatment effects. Throughout we stress randomization-based rather than model-based inference as the basis of understanding inference in randomized experiments.
 We discuss how randomization-based methods relate to more commonly used regression analyses, 
and why we think the emphasis on randomization-based inference is important. We then discuss
the design of experiments, first considering power analyses and then turning to the 
benefits and costs of stratification and pairwise randomization, as well as the complications from re-randomization. 
We recommend using experimental design rather than analysis to adjust for covariates differences in experiments. 
Specifically, we recommend researchers to stratify the population into small strata and then 
randomize within the strata and adjust the standard errors to capture the gains from the stratification.
We argue that this approach is preferred to model-based analyses applied after the randomization 
to adjust for differences in covariates. However, there are limits on how small the strata should be:
we do not recommend to  go as far as pairing the units, because it complicates the analysis due to the fact
that variances cannot be estimated within pairs, whereas they can within strata with at least 
two treated and two control units.
We also discuss in detail methods for estimating heterogenous treatment effects. We focus on methods that allow the researcher to identify subpopulations with different average treatment effects, as well as methods for estimating conditional average treatment effects. In both cases these methods allow the researcher to construct valid confidence intervals.

This chapter draws from a variety of literatures, including the statistical literature on the analysis and design of experiments, e.g., Wu and Hamada (2009), Cox and Reid (2000), Altman (1991), Cook and DeMets (2008), Kempthorne (1952, 1955), Cochran and Cox (1957),  Davies (1954), and Hinkelman and Kempthorne (2005, 2008). We also draw on  the literature on causal inference, both in experimental and observational settings, Rosenbaum (1995, 2002, 2009), Rubin (2006), Cox (1992), Morgan and Winship (2007), Morton Williams (2010) and Lee (2005), and Imbens and Rubin (2015). In the economics literature we build on recent guides to practice in randomized experiments in development economics, e.g., 
Duflo,  Glennerster, and  Kremer (2006), Glennerster (2016),  and Glennerster and   Takavarasha (2013) as well as the general empirical micro literature (Angrist and Pischke, 2008).

There have been a variety of excellent surveys of methodology for experiments in recent years.
Compared to Duflo, Glennerster and Kremer (2006),  Glennerster and Takavarasha (2013) and Glennerster (2016), this chapter focuses more on formal  statistical methods and less on issues of implementation in the field. Compared to the statistics literature, 
we restrict our discussion largely to the case with a single binary treatment. We also pay more attention to the complications arising from non-compliance, clustered randomization, and the presence of interactions and spillovers. Relative to the general causal literature, e.g., Rosenbaum (1995, 2009) and Imbens and Rubin (2015), we do not discuss observational studies with unconfoundedness 
or selection-on-observables in depth, and focus more on complications in experimental settings.

This chapter is organized as follows.  In Section \ref{section:clustering} we discuss the complications arising from cluster-level randomization. 
We discuss how the use of clustering required the researcher to make choices regarding the estimands. 
We also focus on the choice concerning the 
unit of analysis, clusters or lower-level units.  We recommend in general  
to focus on cluster-level analyses as the primary analyses.  Section \ref{section:noncompliance}
 contains a discussion of non-compliance to treatment assignment and its relation to instrumental variables methods.
In Section \ref{section:heterogeneity} we present some recent results for analyzing heterogeneity in treatment effects. 
Finally, Section \ref{section:interactions} we discuss violations of the no-interaction assumption, allowing outcomes for one unit to be affected by treatment assignments for other units. These interactions can take many forms, some through clusters, and some through general networks. We show that it is possible to calculate exact p-values for tests of null hypotheses of no interactions while allowing 
for direct effects of the treatments. Section \ref{section:conclusion} concludes.

\section{Randomized Experiments and Validity}

In this section we discuss some general issues related to the interpretation of analyses of randomized experiments and their validity. 
Following Cochran (1972, 2015) we define randomized experiments as settings where the the assignment mechanism does not depend on characteristics of the units, either observed or unobserved, and the researcher has control over the assignments. In contrast, in observational studies (Rosenbaum, 1995; Imbens and Rubin, 2015), the researcher does not have control over the assignment mechanism, and the assignment mechanism may depend on observed and or unobserved characteristics of the units in the study.
In this section we discuss four specific issues. First, we elaborate on the distinction between randomized experiments and observational studies. Second, we discuss internal validity, and third, external validity. Finally, we discuss the issues related to finite population versus infinite super-population inference.

\subsection{Randomized Experiments versus Observational Studies}
 
There is a long tradition viewing randomized experiments as the most credible of designs to obtain causal  inferences. Freedman (2006) writes succintly ``Experiments offer more reliable evidence on causation than observational studies.''
On the other hand, some researchers continue to be skeptical about the relative merits of randomized experiments. 
For example, Deaton (2010) argues,  that ``I argue that evidence from randomized experiments has no special priority. ... Randomized 
experiments cannot automatically trump other evidence, they do not occupy any special place in some hierarchy of evidence'' (Deaton, p. 426).
Our view aligns with that of Freedman and others who view randomized experiments as playing a special role in causal inference. 
A randomized experiment is unique in the control the researcher has over the assignment mechanism, and by virtue of that control, selection bias in comparisons between treated and control units 
can be eliminated. That does not mean that randomized experiments 
can answer all causal questions. There are a number of reasons 
why randomized experiments may not be suitable to answer particular questions. 

First, consider a case where we are interested in the causal effect of a particular intervention on a single unit: what would the outcome have been for a particular firm in the absence of a merger compared to the outcome given the merger. In that case, and similarly for many questions in macroeconomics,  no randomized experiment will provide us with the answer to the causal question.
Once the interest is in an intervention that can be applied repeatedly, however, it may be possible to conduct experiments, or find data from quasi experiments, even in macroeconomics. Angrist and Kuersteiner  (2011), building on work by Romer and Romer (2004), use the potential outcome framework to discuss causal analyses in a macro-economic time series context.
Second, it may not be ethical to conduct an experiment. In educational settings it is often impossible to withhold particular educational services to individuals in order to evaluate their benefits. In such cases one may need to do observational studies of some kind, possibly randomizing inducements to participate in the programs.

\subsection{Internal Validity}

In a classic text, Shadish, Cook, and Campbell (2002) discuss various aspects of the validity of studies of causal effects. 
Here we focus on two of the most important ones, {internal validity} and {external validity.}
Shadish, Cook, and Campbell (2002) define a study to have  {internal validity} if the observed covariance between a treatment and an outcome reflects ``a causal relationship ... in which the variables were manipulated,'' (p. 53).
Internal validity  refers to the ability of a study to estimate causal effects within the study population. Shadish, Cook, and Campbell (2002) then continue to observe that ``the [internal validity] problem is easily solved in experiments because they force the manipulation  of A to come before the the measurement of B.'' Essentially they argue that well-executed randomized experiments by definition have internal validity, and that the problem of internal validity is one that plagues only observational studies or compromised random experiments. This is not necessarily true in experimental settings where interference between units is a concern.


\subsection{External Validity}

The second aspect of validity that
Shadish, Cook, and Campbell (2002) consider is that of external validity. They write that ``external validity concerns inferences about the extent to which a causal relationship holds over variation in persons, settings, treatments, and outcomes.'' (p 83). Thus, external validity is concerned with generalizing causal inferences, drawn for a particular population and setting, to others, where these alternative settings could involve different populations, different outcomes, or different contexts. 

 Shadish, Cook and Campbell argue for the primacy of internal validity, and claim that without internal validity causal studies have little value. This echos Neyman's comment that without actual randomization a study would have little value, as well as Fisher's observation that randomization was what he called ``the reasoned basis'' for inference. It stands in sharp contrast with a few researchers who have recently claimed that there is no particular priority for  internal validity over external validity (e.g., Manski, 2013). 

The first important point is that external validity cannot be guaranteed, neither in randomized experiments, nor in observational studies. Formally one major reason for that in experiments involving human subjects is that one typically needs informed consent: individuals typically need to agree to participate in the experiment. There is nothing that will guarantee that subjects who agree to do so will be similar to those that do not do so, and thus there is nothing that can guarantee that inferences for populations that give informed consent will generalize to populations that do not. See also the discussion in Glennerster (2016).
This argument has been used to question the value of randomized experiments. However, as Deaton (2010) notes, the same concern holds for nonexperimental studies: ``RCTs, like nonexperimental
results, cannot automatically
be extrapolated outside the context in which
they were obtained'' (Deaton, 2010, p. 449). There is nothing in non-experimental methods that makes them superior to randomized experiments with the same population and sample size in this regard.

Fundamentally, most concerns with external validity are related to treatment effect heterogeneity. 
Suppose one carries out a randomized experiment in setting A, where the setting may be defined in terms of geographic location, or time, or subpopulation. What value have inferences about the causal effect in this location regarding the causal effect in a second location, say setting B. Units in the two settings may differ in observed or unobserved characteristics, or treatments may differ in some aspect. 
To assess these issues it is helpful to have causal studies, preferably randomized experiments, in multiple settings. These settings should vary in terms of the distribution of characteristics of the units, and possibly in terms of the  specific nature of the treatments or the treatment rate, in order to assess the credibility of generalizing to other settings.
An interesting case study is the effect of micro finance programs. Meager (2015) analyzes data from seven randomized experiments, including six published in a special issue of the American Economic Journal (Applied) in 2015, and finds remarkable consistency across these studies.

Another approach is to specifically account for differences in the distributions of characteristics across settings.  Hotz, Imbens, Mortimer (2005) and 
Imbens (2010) set up a theoretical framework where the differences in treatment effects between locations arise from differences in the distributions of characteristics of the units in the locations. 
Adjusting for these differences in unit-level characteristics (by reweighting the units) enables the researcher to compare the treatment effects in different locations.
Allcott (2015) assess the ability of  similar unconfoundedness/selection-on-observable conditions to eliminate differences
in treatment effects between 111 energy conservation programs. Recently developed methods for assessing
treatment effect heterogeneity with respect to observables, reviewed below in Section \ref{section:heterogeneity}, can in principle be used to flexibly 
estimate and conduct inference about treatment effects conditional on observables.

Finally, Bareinboim, Lee, Honavar and Pearl (2013) develop graphical methods to deal with external validity issues.

\subsection{Finite Population versus Random Sample from Super-population}

It is common in empirical analyses to view the sample analyzed as a random sample drawn randomly from a large, essentially infinite super-population. 
Uncertainty is viewed as arising from this sampling, with knowledge of the full population leading to full knowledge of the estimands.
In some cases, however, this is an awkward perspective. In some of these cases the researcher observes all units in the entire 
population, and sampling uncertainty is absent. In other cases it is not clear what population the sample can be viewed as being 
drawn from. 

A key insight is that viewing the statistical problem as one of causal inference allows one to interpret the 
uncertainty as meaningful without any sampling uncertainty. Instead the uncertainty is viewed as arising from the 
unobserved (mising) potential outcomes: we view some  units in the population exposed to one level of the treatment, but do not observe what would have happened to those units had they been exposed to another treatment level, leaving some of the 
components of the estimands unobserved. Abadie, Athey, Imbens and Wooldridge (2014) discuss these issues in detail.

In part of the discussion in this chapter, therefore, we view the sample at hand as the full population of interest, following the approaches taken by Fisher (1925, 1935), Neyman (1935), and subsequently by Rubin (1974, 1978, 2007). The estimands are defined in terms of this finite population. However, these estimands depend on all the potential outcomes, some observed and others not observed, and as a result we cannot infer the exact values of the estimands even if all units in the population are observed. Consider an experiment with ten individuals, five randomly selected to receive a new treatment, and the remaining five assigned to the control group. Even if this group of ten individuals is the entire population of interest, observing realized outcomes for these ten individuals will not allow us to derive the estimand, say the difference in the average outcome if all individuals were treated and the average outcome if all ten individual were to receive the control treatment, without uncertainty. The uncertainty is coming from the fact that for each individual we can only see one of the two relevant outcomes.
In many cases the variances associated with estimators based on random assignment of the treatment will be similar to those calculated conventionally based on sampling uncertainty. In other cases the conventional sampling-based standard errors will be unnecessarily conservative. When covariates are close to uncorrelated with the treatment assignment (as in a randomized experiment), the
differences are likely to be modest.  See Abadie, Athey, Imbens and Wooldridge (2014) for details.

\section{The Potential Outcome  / Rubin Causal Model Framework for Causal Inference}
\label{section:causality}

The perspective on causality we take in this chapter is associated with the {potential outcome} framework (for a textbook discussion see Imbens and Rubin, 2015). This approach goes back to Fisher (1925) and Neyman (1928). The work by Rubin (1973, 1975, 1978) led Holland (1986) to label it the Rubin Causal Model (RCM). 

\subsection{Potential Outcomes}

This RCM  or potential outcome setup has three key features. The first is that it associates causal effects with potential outcomes. For example, in a setting with a single unit (say an individual), and a single binary treatment, say taking a drug or not, we associate two potential outcomes with this individual, one given the drug and one without the drug. The causal effect is the comparison between these two potential outcomes. The problem, and in fact what Holland (1985) in a widely quoted phrase called the ``fundamental problem of causal inference'' (Holland, 1986, p. 947) is that we can observe at most one of these potential outcomes, the one corresponding to the treatment received. In order for these potential outcomes to be well defined, we need to be able to think of a manipulation that would have made it possible to observe the potential outcome that corresponds to the treatment that was not received, which led Rubin to claim ``no causation without manipulation'' (Rubin, 1975, p. 238). Because for any single unit we can observe at most one of the potential outcomes, we need to observe outcomes for multiple units. This is the second feature of the potential outcomes framework, the necessity of the presence of multiple units. By itself the presence of multiple units does not solve the problem because with multiple units the number of distinct treatments increases: with $N$ units and two treatment levels for each unit there are $2^N$ different values for the full vector of treatments, with any comparison between two of them a valid causal effect.  However, in many cases we are willing to make assumptions that interactions between units is limited so that we can draw causal inferences from comparisons between units. An extreme version of this is the assumption that the treatment for one unit does not affect outcomes for any other unit, so that there is no interference whatsoever. The third key feature of the RCM is the central role of the assignment mechanism. Why did a unit receive the treatment it did receive? Here randomized experiments occupy a special place in the spectrum of causal studies: in a randomized experiment the assignment mechanism is a known function of observed characteristics of the units in the study. The alternative, where parts of the assignment mechanism are unknown, and may possibly depend on unobserved characteristics (including the potential outcomes) of the units, are referred to as { observational studies} (Cochran, 1972).

There are alternative approaches to causality. Most notably there has been much work recently on {causal graphs}, summarized in the book by Pearl (2000, 2009). In this approach causal link are represented by arrows and conditional independencies are captured by the absence of arrows. These methods have been found useful in studies of identification questions as well as for using data to discover
causal relationships among different variables. However, the claims in this literature (e.g., Pearl, 2000, 2009) that the concept of a causal effect does not require the ability to at least conceptually manipulate treatments remains controversial.

Let us now add some specifics to this discussion. Suppose we start with a single unit, say ``I''. Suppose we have a binary treatment, denoted by $W\in\{0,1\}$ for this unit, which may correspond to taking a drug or not. The two potential outcomes are $Y(0)$, the outcome for me if I do not take the drug, and $Y(1)$, the outcome if I do take the drug. The causal effect is a comparison of the two potential outcomes, say the difference, $Y(1)-Y(0)$, or the ratio, $Y(1)/Y(0)$. Once we assign the treatment, one of the potential outcomes will be realized and possibly observed:
\[ Y^\obs=Y(W)=
\left\{
\begin{array}{ll} Y(0)\hskip1cm & {\rm if}\  W=0,\\
 Y(1)\hskip1cm & {\rm if}\  W=1\end{array}\right.
\]
We can only observe one of the potential outcomes, so drawing credible and precise inferences about the causal effect, say the difference $Y(1)-Y(0)$ is impossible without additional assumptions or information. Now let us generalize to the setting with $N$ units, indexed by $i=1,\ldots,N$. Each of the units can be exposed to the two treatments, no drug or drug, with $W_i$ denoting the treatment received for unit $i$.
 Let $\bw$ be the $N-$vector of assignments with typical element $W_i$. The problem is that in principle the potential outcomes can depend on the treatments for all units, so that for each unit we have $2^N$ different potential outcomes $Y_i(\bw)$. In many cases it is reasonable to assume that the potential outcomes for unit $i$ depend solely on the treatment received by unit $i$. This is an important restriction on the potential outcomes, and one that is unrealistic in many settings, with a classic example being that of vaccinations for  infectious diseases. For example, exposing some students to educational interventions may affect outcomes for their class mates, or training some unemployed individuals may affect the labor market prospects for other individuals in the labor market. We will discuss the complications arising from interactions in Section \ref{section:interactions}. Note that the interactions can be a nuisance for estimating the effects of interest, but they can also be the main focus.

If we are willing to make the no-interference assumption, or sutva (stable unit treatment value assumption, Rubin, 1978), we can index the potential outcomes by the own treatment only, and write without ambiguity $Y_i(w)$, for $w=0,1$. For each of the $N$ units the realized outcome is now $Y^\obs_i=Y_i(W_i)$. Now with some units exposed to the active treatment and some exposed to the control treatment, there is some hope for drawing causal inferences. In order to do so we need to make assumptions about the assignment mechanism. To be formal, let $\mmy$ be the range of values for the potential outcomes, and let $\mmx$ be the range of values for the covariates or pretreatment variables. In general we write this as a function
\[ p:\{0,1\}^N\times \mmy^{2N}\times \mmx^N\mapsto [0,1],
\]
so that $p(\bw|\by(0),\by(1),\bx)$ is the probability of the assignment vector $\bw$, as a function of all the potential outcomes and covariates. 

We limit the general class of assignment mechanism we consider. The most important limitation is that for randomized experiments we disallow dependence on the potential outcomes, and we assume that the functional form of the assignment mechanism is known.
Analyzing observational studies where the assignment mechanism depends in potentially complicated ways on the potential outcomes is often a challenging task, typically relying on controversial assumptions.

\subsection{A Classification of Assignment Mechanisms}

Let us consider four assignment mechanisms that we will discuss in subsequent sections in this chapter.

\subsubsection{Completely Randomized Experiments}
 In { completely randomized experiment} a fixed number of units, say $\nt$, is drawn at random from the population of $N$ units to receive the active treatment, with the remaining $\nc=N-\nt$ assigned to the control group. 
It satisfies
\[p(\bw|\by(0),\by(1),\bx)=\left(
\begin{array}{c}N \\ \nt\end{array}\right)^{-1}\hskip1cm
{\rm for\ all}\  \bw\ {\rm such\ that} \sum_{i=1}^N W_i=\nt.\] 

\subsubsection{Stratified Randomized Experiments}
\label{stratification}

The next two experimental designs, stratification and pairing, are intended to improve the efficiency of the design by disallowing assignments that are likely to be uninformative about the treatment effects of interest.
In a { stratified randomized experiment} we first partition the population on the basis of covariate values into $G$ strata. Formally, if the covariate space is $\mmx$, we partition $\mmx$ into $\mmx_1,\ldots,\mmx_G$, so that $\cup_g\mmx_g=\mmx$, and $\mmx_j\cap\mmx_{g'}=\emptyset$ if $g\neq g'$. Let $G_{ig}$ be an indicator for unit $i$ belonging to stratum $g$, so that $G_{ig}={\bf 1}_{X_i\in\mmx_g}$. Let $N_g$ be the number of units in stratum $g$. Then we fix the number of treated units in each stratum as $N_{{\rm t},g}$, so that the total number of treated units is $\nt=\sum_{g=1}^G N_{{\rm t},g}$. The assignment probability is then
\[p(\bw|\by(0),\by(1),\bx)=\prod_{g=1}^G\left(
\begin{array}{c}N_j \\ N_{{\rm t},g}\end{array}\right)^{-1},\hskip0.5cm
{\rm for\ all}\  \bw\ {\rm such\ that}\ \forall g \ \sum_{i=1}^N G_{ij}\cdot W_i=N_{{\rm t},g}.\] 
This design rules out some assignments that are allowed in a completely randomized design, with the hope that the assignment vectors that are disallowed are relatively uninformative compared to assignment vectors that are allowed, for example where all men are in the treatment group and all women in the control group.

\subsubsection{Paired Randomized Experiments}

In a {paired randomized experiment} we pair units together and randomize within the pairs. We can think of this as an extreme case of stratification where each stratum contains exactly one treated unit and exactly one control unit. In that case there are $G=N/2$ strata, and 
$N_g=2$ and $N_{{\rm t},g}=1$ for all $g$. Then
\[p(\bw|\by(0),\by(1),\bx)=
\left(\frac{1}{2}\right)^{N/2}\hskip1cm
{\rm for\ all}\  \bw\ {\rm such\ that}\ \forall g, \ \sum_{i=1}^N G_{ig}\cdot W_i=1.\] 

\subsubsection{Clustered Randomized Experiments}

The last design we discuss is not  intended to be more informative than a completely randomized experiment with the same sample size. Rather it is a design that attemps to avoid 
complications with local interactions at the unit level,
as well as disallow assignments that are may be relatively expensive in terms of data collection, and thus indirectly may  attempt to increase the sample size to improve precision.
In a { clustered randomized experiment}, as in a stratified randomized experiments, we start with a partitioning of the covariate space. Now, however, instead of assigning treatments randomly to units within a cluster (the same as the stratum in the stratified randomized experiment), treatments are assigned randomly to entire clusters, with all units within a cluster receiving the same level of the treatment. 

This design may be motivated by concerns that there are interactions between units. For example, for educational programs it may be that exposing some children in a classroom to an intervention has spillover effects on children in the same classroom who were not exposed to the intervention. For that reason is may make sense to expose all children in a classroom or school to the same treatment. Alternatively, it may be expensive to randomize at the individual level compared to randomizing at the classroom or geographic unit level.

Again, let $G_{ig}$ denote the indicator for unit $i$ belonging to cluster $g$, with $G$ the total number of clusters. Although we may vary the probability of a cluster being assigned to the treatment group, here we focus on the simplest case where $G_\ttt$ out of the $G$ clusters are selected randomly to be assigned to the treatment group. Thus, at the cluster level we have a completely randomized experiment.
Let $\overline{W}_g=\sum_{i:G_{ig}=1} W_i/N_g$ be the average value of $W_i$ for units in cluster $g$, so that the clustering implies that $\overline{W}_g\in\{0,1\}$.
More generally, one may vary the probability of being assigned to the treatment by cluster, without requiring that all units in the same cluster having the same treatment, although we do not consider that case here.
Then:
\[p(\bw|\by(0),\by(1),\bx)=\left(
\begin{array}{c}G \\ G_\ttt\end{array}\right)^{-1},
\] 
for all $\bw$   such that if $G_{ig}=G_{i'g}=1$,  then $W_i=W_{i'},$ and $\sum_{g=1}^G \overline{W}_g=G_\ttt.$

\section{The Analysis of Completely Randomized Experiments}
\label{section:randomizedexperiments}

In this section we discuss the analysis of the simplest form of randomized experiments, completely randomized experiments. In this setting we have a sample of $N$ units, $\nt$ of whom are selected at random to receive the active treatment, and the remaining $\nc=N-\nt$ of whom receive the control treatment. We consider four sets of analyses. First, we study the calculation of exact p-values for sharp hypotheses, based on the work by Fisher (1925, 1935). Second, we consider estimation of and inference for average treatment effects, following the original work by Neyman (1928, 1935, 1990). Third, we study the relation between the Neyman approach and linear regression, showing how randomization justifies conventional regression analyses. Fourth, we look at quantile treatment effects. Central to our discussion is the view of the potential outcomes as fixed, leading to a focus on inference based on the randomization distribution, keeping fixed the total number of units assigned to treatment and control. We will sometimes view the sample as identical to the population of interest, and sometimes as a random sample from an infinitely sized population of interest.

Initially we focus on the case without pretreatment variables. 
In Section \ref{ran_cov} we allow for the presence of covariates but maintain the focus on global targets such as the average effect of the treatment.
In Section \ref{section:heterogeneity} we explore the benefits of observing covariates that are not affected by the treatment, also known as pre-treatment variables.
We will illustrate some of the discussions with analyses of an experimental evaluation of a labor market program, first analyzed by Lalonde (1986). The data set contains information on 445 individuals, 185 in the treatment group, and 260 in the control group. The outcome is post training earnings, and pre-treatment variables include  lagged earnings and individual characteristics.

\subsection{Exact P-values for Sharp Null Hypotheses}
\label{section:fisher}

The first analysis is based on Fisher's (1925, 1935) work on exact p-values for sharp null hypotheses.
See for recent discussions Rosenbaum (1992), Gail, Tian and Piantadosi (1988), and Imbens and Rubin (2015), and in economics  Young (2016). 
Fisher was interested in testing sharp null hypotheses, that is, null hypotheses under which we can infer all the missing potential outcomes from the observed ones. The leading null hypothesis in this class is the null hypothesis that the treatment has no effect whatsoever:
\begin{equation} H_0:\ \ Y_i(0)=Y_i(1)\ \ \ \forall\ i=1,\ldots,N.\end{equation}
The implicit alternative hypothesis is that there is at least one unit $i$ such that $Y_i(0)\neq Y_i(1)$. Other sharp null hypothesis correspond to known constant treatment effects, but in many cases these are less interesting and natural. However, in some cases
one can use the exact p-values in settings without sharp null hypotheses by redefining the experiment, as shown by Athey, Eckles and Imbens (2015) in the context of network experiments (see Section \ref{section:networks} for further discussion).

Given the sharp null hypothesis, we can infer all the missing potential outcomes. As a result we can infer, for any statistic that is a function of $\by^\obs$, $\bw$, and $\bx$, the exact distribution of that statistic under the null hypothesis. So, suppose we choose as our statistic the difference in means by treatment status:
\begin{equation} T^\ave(\bw,\by^\obs,\bx)=
\oy^\obs_\ttt-\oy^\obs_\ccc
=\frac{1}{\nt}\sum_{i:W_i=1} Y^\obs_i-\frac{1}{\nc}\sum_{i:W_i=0} Y^\obs_i.
\end{equation}
We can calculate the probability, over the randomization distribution, of the statistic taking on a value as large, in absolute value, as the actual value given the actual treatment assigned. This calculation gives us the p-value for this particular null hypothesis:
\begin{equation}p= {\rm pr}\left(|T^\ave(\bw,\by^\obs,\bx)|\geq |T^\ave(\bw^\obs,\by^\obs,\bx)|\right).\end{equation}

Let us illustrate this using data from National Supported Work program, previously analyzed by Lalonde (1986), Dehejia and Wahba (1999) and many others.
The simple difference in average post treatment earnings between treated and control is 1.79 (in thousands of dollars).
To calculate the p-value associated with this difference of 1.79, we re-assign the treatment, keeping the number of treated and control units fixed at 185 and 240 respectively. Given the reassigned treatment we calculate what the value of the statistic would have been. Although the observed outcomes do not change for any unit under the null hypothesis, the value of the statistic changes because who is in the treatment group and who is in the control group changes. Repeating this many times we calculate the fraction of reassignment vectors that leads to a statistic that is at least as large as 1.79 in absolute value.
 The p-value associated with this statistic is 0.0044, suggesting we should clearly reject the null hypothesis that the program had no effect on earnings whatsoever.

The main choice to be made in this procedure is the choice of statistic. A natural statistic is the one we choose in the illustration, the difference in means by treatment status. Another attractive choice is the difference in means of the ranks by treatment status. Here the outcomes are first converted to ranks, normalized to have zero mean:
\[R_i=R(i;Y_1^{\rm obs},\ldots,Y_N^{\rm obs})=\sum_{j=1}^N
\been_{Y_j^{\rm obs}<Y_i^{\rm obs}}
+\frac{1}{2}\left(1+\sum_{j=1}^N \been_{Y_j^{\rm obs}=Y_i^{\rm obs}}\right)
-\frac{N+1}{2}.
\]
The term in the middle deals with the presence of ties in the data.
For the Lalonde data this statistic leads to a p-value of 0.01. In this application the robustness to outliers does not actually buy very much, and the presence of many zeros has a bigger impact on the difference between the mean and rank statistics.

This transformation improves the power of the tests in settings with outliers and thick-tailed distributions. It is less arbitrary than, for example, simply transforming the outcome by taking logarithms, especially in settings where such transformations are not feasible, e.g., in settings with thicktailed distribution and a mass point at zero. There are some settings where the transformation to ranks does not work well. An example would be a case with a large proportion of zeros, and a very thick-tailed distribution for the outcomes for units with non-zero outcomes.

In some cases the researcher has multiple outcomes. One can calculate exact p-values for each of the outcomes, but obviously the probability that at least one of the p-values is less than 0.05 even if the treatment has no effect on any of the outcomes is generally larger than 0.05. There are two modifications one can implement to address this. The simplest is to modify the test statistic to take account of all the outcomes. For example, one could use an F-statistic, that is, a quadratic form in the difference in average outcomes by treatment status, with the inverse of the covariance matrix in the middle. For that statistic one can calculate the exact p-value under the null hypothesis that there is no effect of the treatment whatsoever using the Fisher randomization distribution. See for example Young (2016). Alternatively one can use adjustments to the p-values to take account of the multiple testing. Traditionally such adjustments are based on Bonferroni bounds. However, there are tighter bounds available, although they tend to be more conservative than the exact Fisher p-values. See Romano, Shaikh and Wolf (2010) for a review of this literature.

Rosenbaum (1992) discusses estimators for treatment effects based on rank statistics, as opposed to simply doing tests, following Hodges and Lehmann (1970) and Doksum (1974). Specifically he looks for values for the common treatment effect that set the rank correlation between the residuals and the treatment equal to zero, leading to confidence intervals based on inverting test statistics.

\subsection{Randomization Inference for Average Treatment Effects}
\label{section:neyman}

In this section we continue the analysis of completely randomized experiments, taking as fixed the potential outcomes in the population. Here we follow the line of research that originates in the work by Neyman (1929, 1935, 1990). Neyman was interested in estimating the average effect of the treatment for the sample at hand,
\begin{equation}
\tau=\frac{1}{N}\sum_{i=1}^N \Bigl( Y_i(1)-Y_i(0)\Bigr)=\oy(\ct)-\oy(\cc).
\end{equation}
 In addition Neyman was interested in constructing confidence intervals for such average effects. Initially we focus purely on the finite sample, with no assumptions on any sampling that may have led to the particular sample at hand.

As an estimator Neyman proposed the difference in average outcomes by treatment status:
\begin{equation} \hat\tau=\oy_\ttt^\obs-\oy^\obs_\ccc, \hskip1cm {\rm where}\ \ 
\oy_\ttt^\obs=\frac{1}{N_{\rm t}}\sum_{i:W_i=1} Y^\obs_i,\ \ {\rm and}\ \ 
\oy_\ccc^\obs=\frac{1}{N_{\rm c}}\sum_{i:W_i=0} Y^\obs_i
.\end{equation}
Defining 
\[ D_i=W_i-\frac{\nt}{N}
=\left\{\begin{array}{ll}
\frac{\nc}{N}\hskip1cm &\rif\ W_i=\ct\\
-\frac{\nt}{N} & \rif\ W_i=\cc,\end{array}
\right.
\] 
so that $\mme[D_i]=0$,
we can write this  estimator as: 
\begin{equation}\label{d}\hat\tau=\tau
+\fen\sumn D_i\cdot 
\left(\frac{N}{\nt}\cdot \yie+\frac{N}{\nc}\cdot\yin\right).
\end{equation}
Because  all potential outcomes are fixed, the only stochastic components are the $D_i$,  and with $\mme[D_i]=0$, 
the second term has expectation zero, which immediately implies that this estimator 
is unbiased for the
average treatment effect, $\tau$. 
A more tedious calculation (e.g., Imbens and Rubin, 2015), shows that the sampling variance of $\hat\tau$, over the randomization distribution, is:
\begin{equation}
\mmv\left(\hat\tau\right)=
\frac{\sscs}{\nc}+\frac{\ssts}{\nt}-\frac{S_{\ttt\tc}^2}{N},
\label{vareen}
\end{equation}
where $\sscs$ and $\ssts$ are the variances of $\yin$ and $\yie$ in the sample,
defined as:
\[S_\tc^2=\frac{1}{N-1}\sum_{i=1}^{N}\Bigl(\yin-\oy(\cc)\Bigr)^2,\aand 
S_\ttt^2=\frac{1}{N-1}\sum_{i=1}^{N}\Bigl(\yie-\oy(\ct)\Bigr)^2,
\]
and $S_{\ttt\tc}^2$ is the sample variance of the unit-level
treatment effects, defined as:
\[S_{\ttt\tc}^2=\frac{1}{N-1}\sum_{i=1}^{N}
\Bigl(\yie-\yin-
(\oy(\ct)-\oy(\cc))\Bigr)^2.\]
We can estimate the first two terms as
\[ \scs
=\frac{1}{\nc-1}\sum_{\iwc} \left(\yin
-\oyobsn\right)^2=\frac{1}{\nc-1}\sum_{\iwc}  \left(\yoi
-\oyobsn\right)^2 ,\]
and
\[ \sts=\frac{1}{\nt-1}\sum_{\iwt} \left(Y_i(\ct)
-\oyobse\right)^2=\frac{1}{\nt-1}\sum_{\iwt}\left(\yoi
-\oyobse\right)^2.\]
These estimators are unbiased for the corresponding terms in the variance of $\hat\tau$.
The third term, $S_{\ttt\tc}^2$
(the population variance of the unit-level treatment
effects $Y_i(1)-Y_i(0)$) is generally impossible
to estimate consistently because we  never observe both $\yie$ and $\yin$
for the same unit. We therefore have no direct
observations on the variation in the treatment effects across the population
and  cannot directly estimate $S_{\ttt\tc}^2$.

In practice researchers therefore use the estimator for $\mmv\left(\hat\tau\right)$ based on estimating the first two terms by $\scs$ and $\sts$, and ignoring the third term,
\begin{equation}
\hat{\mmv}_\neyman=
\frac{\scs}{\nc}+\frac{\sts}{\nt}.
\label{vareen1}
\end{equation}
 This leads in general to an upwardly biased estimator for $\mmv\left(\hat\tau\right)$, and thus to conservative confidence intervals. 
There are two important cases where the bias vanishes. First, if the treatment effect is constant the third term is zero, and so ignoring it is immaterial. Second, if we view the sample at hand as a random sample from an infinite population, then $\mmv\left(\hat\tau\right)$ is unbiased for the variance of $\hat\tau$ viewed as an estimator of the population average treatment effect $\mme[Y_i(1)-Y_i(0)]$, rather than as an estimator of the sample average treatment effect
$\sum_{i=1}^N (Y_i(1)-Y_i(0))/N$ (See Imbens and Rubin, 2015).

To construct confidence intervals we do need to make large sample approximations. One way to do this is to assume that the sample can be viewed as a random sample from a large population and use a standard central limit theorem for independent and identically distributed random variables. An alternative is to make assumptions on the properties of the sequence of $(Y_i(0),Y_i(1))$ so that one can use a Lindenberg-type central limit theorem for independent, but not identically distributed, random variables for the second term in (\ref{d}). The main condition is that the sequence of averages of the  squares of $Y_i(0)+Y_i(1)$ does not diverge. The large sample approximations do play a very different role though than in standard discussions with random sampling. Most importantly the estimand is defined in terms of the finite sample, not in terms of the infinite superpopulation.

For the Lalonde data the estimate and Neyman standard error, which are up to a degrees-of-freedom adjustment equal to the White robust standard errors (White, 1980), are
\[ \hat\tau= 1.794\ \ (\widehat{\rm se}\    0.671).\] 
The p-value based on the normal approximation to the distribution of the t-statistic is 0.0076, compared to an exact p-value of 0.0044 based on the Fisher approach.

\subsection{Quantile Treatment Effects}
\label{section:quantile}

Much of the theoretical as well as the empirical  literature on treatment effects has focused on average causal effects. However, there are other causal effects that might be of interest. Of particular interest are quantile treatment effects. 
These can be used as a systematic way to uncover treatment effects that may be concentrated in tails of the distribution of outcomes, or to estimate more robustly constant treatment effects in settings with thick-tailed distributions. For this case there are no finite sample results in the spirit of Neyman's results for the average treatment effect, so we focus on the case where the sample can be viewed as a random sample from an infinite population.

In general, let $q_Y(s)$ denote the $s$-th quantile of the distribution of the random variable $Y$. Formally,
\[  q_Y(s)=\inf_{y} {\bf 1}_{{F_Y(y)\geq s}}.\]
Now define the $s$-th quantile treatment effect as the difference in quantiles between the $Y_i(1)$ and $Y_i(0)$ distributions:
\begin{equation} \tau_s=q_{Y(1)}(s)-q_{Y(0)}(s).\end{equation}
Such quantile treatment  effects have been studied in
 Doksum (1974) and Lehman (1974), and more recently in Abadie, Angrist and Imbens (2002), Chernozhukov and Hansen (2005), Firpo (2007), Bitler, Gelbach and Hoynes (2002).

Note that $\tau_s$ is a difference in quantiles, and  in general it is different from the quantile of the differences, that is, the corresponding quantile of the unit-level treatment effects,
$q_{Y(1)-Y(0)}(s)$. Specifically, although the mean of the difference of $Y_i(0)$ and $Y_i(0)$ is equal to the difference in the means of $Y_i(1)$ and $Y_i(0)$,  in general  the median of the difference $Y_i(1)-Y_i(0)$ is not equal to the difference in the medians of $Y_i(1)$ and $Y_i(0)$.
There are three important issues concerning the quantile of the treatment effects in relation to the differences in quantiles.
First, 
the two estimands, $q_{Y(1)}(s)-q_{Y(0)}(s)$ and 
$q_{Y(1)-Y(0)}(s)$, are equal if there is perfect rank correlation between the two potential outcomes. In that case,
\[ Y_i(1)=F^{-1}_{Y(1)}(F_{Y(0)}(Y_i(0))).\]
A special case of this is that where the treatment effect is additive and constant.
This assumption is implausible in many settings. However, in general it has no testable implications.

The second, related, issue is that in general the
quantile of the unit-level treatment effects,
$q_{Y(1)-Y(0)}(s)$, is not identified. Even with large scale experiments we can only infer the two marginal distributions of $Y_i(0)$ and $Y_i(1)$. Nothing about the joint distribution that  cannot be expressed in terms of these two marginal distributions can be inferred from the data.

A third issue is the question which of the two quantile treatment effects, 
the difference in quantiles, $q_{Y(1)}(s)-q_{Y(0)}(s)$, or the quantile of the difference, 
$q_{Y(1)-Y(0)}(s)$, is the more interesting object for policy makers. To discuss that question it is useful to think about the possible decisions
 faced by a policy maker. If a policy maker is commited to making one of the two treatments universal and is deciding between exposing all units to the control treatment  or to the active treatment, the answer should depend only on the  two marginal distributions, and not on aspects of  the joint distribution that cannot be expressed in terms of the two marginal distributions. This suggests that the difference in quantiles may be a more natural object to consider, although there are some cases, such as legal settings, where unit-level treatment effects are of primary interest.

For these reasons we focus on the difference in quantiles, $\tau_s$. Inspecting this estimand for different values of $s$ may reveal that a particular treatment affects the lower or upper tail more than the center of the distribution. In addition, in cases where the average effects of the treatment may be imprecisely estimated because of thick-tailed distributions, quantile treatment effect estimates may be very informative.

Here we estimate quantile effects for the Lalonde data for the quantiles 0.10, 0.25, 0.50, and 0.75. For each quantile we estimate the average effect and calculate standard errors using the bootstrap. We also use the difference in quantiles as a statistic in an exact p-value calculation.
 \begin{table}[ht]
 \caption{\sc  Estimates of Quantile Treatment Effects for Lalonde Data.}
 \vskip1cm
 \begin{center}
 \begin{tabular}{cccc}
 \\
quantile & est & bootstrap s.e. & exact p-value\\
\\
0.10 & 0.00 & (0.00)& 1.000\\
0.25 &   0.49 &    (0.35)& 0.003\\
0.50&1.04 & (0.90)&0.189 \\
0.75 &2.34 &(0.91)&  0.029 \\
0.90 &2.78&(1.97)&  0.071\\
 \end{tabular}
 \end{center}
 \label{tabel1}
\end{table}
The results for the exact tests are quite different from those based on estimating the effects and calculating standard errors. The reason is that the quantile estimates are far from normally distributed. Mainly because of the 30\% zeros in the outcome distribution the distribution of the difference in the lower quantiles has a substantial point mass at zero. Because of the substantial proportion of individuals with zero earnings, the bootstrap standard error for the 0.10 quantile is essentially zero.

\subsection{Covariates in Completely Randomized Experiments}
\label{ran_cov}

In this section we discuss some additional analyses that a researcher may wish to carry out if covariates are recorded for each unit. Later we discuss regression methods, but here we discuss some general principles. 
We focus here on the case where the randomization took place without taking into account the covariates. In fact, as we discuss in Section \ref{stratification}, if one has covariates observed prior to the randomization, one should modify the design of the experiment and carry out a stratified randomized experiment rather than as a completely randomized experiment. 
If one has a well-conducted randomized experiment where the randomization did not take into account the covariates,  one does not need regressors in order to estimate average treatment effects. The simple difference in means by treatment status, $\hat\tau=\oy_\ttt^\obs-\oy^\obs_\ccc$ is unbiased for the average effect. So, the question is what the role is of covariates. There are  two principal roles. 

First, incorporating covariates may make analyses more informative. For example, one can construct test statistics in the Fisher exact p-value approach that may have more power than statistics that do not depend on the covariates. Similary, by estimating average treatment effects within subpopulations, and then averaging up the estimates appropriately, the results will be more precise if the covariates are sufficiently strongly correlated with the potential outcomes. There is potentially a small-sample cost to ex post adjustment.  For example, if the covariates are  independent of the potential outcomes, this ex post adjustment will lower precision slightly. 
In practice the gains in precision tend to be modest.

Second, if the randomization was compromised, adjusting for covariate differences may remove biases. Even if the original randomization was done appropriately, this may be  relevant if there are missing data and the analysis uses only complete cases where there is no guarantee of ex ante comparability between treated and control units.

To illustrate this, let us consider the Lalonde data, and focus on the indicator that the lagged earnings are positive as a covariate.
The overall estimate of the average treatment effect is
\[ \hat\tau=    1.79 \hskip0.5cm (\widehat{\rm se}=0.67).\]
For the individuals with positive prior earnings the effect is
\[ \hat\tau_p=    1.69 \hskip0.5cm (\widehat{\rm se}_p=1.31).\]
For the individuals with zero prior earnings the effect is
\[ \hat\tau_z=    1.71 \hskip0.5cm (\widehat{\rm se}_z=0.74).\]
Combining the two estimates leads to
\[ \hat\tau=\hat p\cdot\hat\tau_p+(1-\hat p)\cdot \hat\tau_z=    1.70 \hskip0.5cm (\widehat{\rm se}=0.66),\]
with a  standard error that is barely smaller than that without adjusting for positive prior earnings, 0.67.

The two arguments regarding the role of covariates in the analysis of randomized experiments also raise the question whether there is any reason to compare the covariate distributions by treatment status as part of the analysis. There are a couple of reasons why such a comparison may be useful. If there is some distance between the agencies carrying out the original randomization and the researcher analyzing the data it may be useful as check on the validity of the randomization to assess whether there are any differences in covariates. Second, even the randomization was carried out appropriately, it may be informative to see whether any of the key covariates were by chance relatively imbalanced between treatment and control group, so that prior to seeing the outcome data an analysis can be designed that addresses these presumably modest imbalances. Third, if there is reason to believe that the sample to be analyzed is not identical to the population that was randomized, possibly because of attrition, or item non-response with incomplete observations dropped from the analysis, it is useful to assess how big the imbalances are that resulted from the sample selection. Tabel \ref{tabel_lal_cov} presents the differences in covariates for the experimental Lalonde data.

\begin{table}[ht]
 \caption{\sc Covariates in Lalonde Data}
 \vskip1cm
 \begin{center}
 \begin{tabular}{lccccc}
 \\
 & \multicolumn{2}{c}{Average} &  & \\
Covariate & Treated & Controls & Difference & s.e. & exact p-value\\
\\
African-American & 0.84 & 0.83 & 0.02 & (0.04) & 0.700 \\  
Hispanic  & 0.06 & 0.11 & -0.05 & (0.03) & 0.089 \\ 
age & 25.8 & 25.0 & 0.8 & (0.7) & 0.268 \\
education & 10.3 & 10.1 & 0.3 & (0.2) & 0.139 \\
married & 0.19 & 0.15 & 0.045 & (0.04) & 0.368 \\  
no-degree & 0.71 & 0.84 & -0.13 & (0.04) & 0.002 \\ 
earnings 1974 & 2.10 & 2.11 & -0.01 & (0.50) & 0.983 \\ 
unemployed 1974 & 0.71 & 0.75 & -0.04 & (0.04) & 0.329 \\ 
earnings 1974 & 1.53 & 1.27 & 0.27 & (0.31) & 0.387 \\  
unemployed 1975 & 0.60 & 0.69 & -0.09 & (0.05) & 0.069 \\ 
 \end{tabular}
 \end{center}
 \label{tabel_lal_cov}
\end{table}
We see that despite the randomization there is substantial evidence that the proportion of individuals with a degree in the treatment group is lower than in in the control group. This conclusion survives adjusting for the multiplicity of testing.

\section{Randomization Inference and Regression Estimators}
\label{section:regression}

In this section we discuss regression and more generally modelling approaches to estimation and inference in the context of completely randomized experiments. Although these methods remain the most popular way of analyzing data from randomized experiments, we suggest caution in using them. Some of these comments echo the concerns raised by others. For example, in the abstract of Freedman (2008), he writes ``Regression adjustments are often made to experimental data. Since randomization does not justify the
models, almost anything can happen'' (Freedman, 2008, abstract) and similar comments are made by Deaton (2010)  Young (2016), and Imbens and Rubin (2015).
Regression methods were not originally developed for analyzing data from randomized experiments, and the attempts to fit the appropriate analyses into the regression framework requires some subtleties.
In particular there is a  disconnect between  the way the conventional assumptions in regression analyses are formulated and the  implications of randomization. As a result it is easy for the researcher using regression methods to go beyond analyses that are justified by randomization, and end up with analyses that rely on a difficult-to-assess mix of randomization assumptions, modelling assumptions, and large sample approximations. This is particularly true once one uses nonlinear methods. See for additional discussions
Lesaffre and Senn (2003), Samii and Aronow (2012), Rosenbaum (2002), Lin (2013), Schochet (2010), Young (2016), Senn (1994).

Ultimately we recommend that researchers wishing to use regression or other model-based methods rather than the randomization-based methods we prefer, do so with care. For example, using only indicator variables based on partitioning the covariate space, rather than using multi-valued variables as covariates in the regression function  preserves many of the finite sample properties that simple comparisons of means have, and leads to regression estimates with clear interpretations.
In addition, in many cases the potential gains from regression adjustment can also be captured by careful ex ante design, that is,  through stratified randomized experiments to be discussed in the next section, without the potential costs associated with ex post regression adjustment.


\subsection{Regression Estimators for Average Treatment Effects}
\label{section:regression}

In  ordinary least squares, one regresses the observed outcome $Y^{\obs}_i$ on the indicator for the treatment, $W_i$, and a constant:
\begin{equation} Y_i^{\obs}=\alpha+\tau\cdot W_i+\varepsilon_i,\label{regression}\end{equation}
where $\varepsilon_i$ is an unobserved error term.
The least squares estimator for $\tau$ is
based on minimizing the sum of squared residuals over $\alpha$ and $\tau$,
\[ (\hat\tau_{\ols},\hat\alpha_{\ols})=\arg\min_{\tau,\alpha}\sum_{i=1}^N (Y^{\rm obs}_i-\alpha-\tau\cdot W_i)^2,\]
with solution
\[ \hat\tau_{\ols}=\frac{\sum_{i=1}^N (W_i-\overline{W})\cdot(Y^{\obs}_i-\oy^{\obs})}{\sum_{i=1}^N (W_i-\overline{W})^2}=\oy_\ttt^\obs-\oy^\obs_\ccc,\hskip1cm
{\rm and}\ \ 
\hat\alpha_{\ols}=\oy^{\obs}-\hat\tau_{\ols}\cdot\overline{W}.\]
The least squares estimate of $\tau$ is identical to the simple difference in means, so by the Neyman results discussed in Section \ref{section:neyman}, the least squares estimator is unbiased for the average causal effect. However, the assumptions that are typically used to justify linear regression are substantially different from the randomization that justifies Neyman's analysis. In addition, the unbiasedness claim in the Neyman analysis is conceptually different from the one in conventional regression analysis: in the first case the  repeated sampling paradigm keeps the potential outcomes fixed and varies the assignments, whereas in the latter the realized outcomes and assignments are fixed but different units with different residuals, but the same treatment status, are sampled.
The assumptions typically used in regression analyses are that, in the infinite population the sample was drawn from, the
error terms $\varepsilon_i$
are independent of,
or at least uncorrelated with, the treatment indicator $W_i$.
This assumption is  difficult to evaluate,
 as the interpretation of these residuals is
rarely made explicit beyond a vague notion of capturing unobserved
factors affecting
the outcomes of interest.
Textbooks therefore often stress that regression estimates measure
 only association between the two variables, and that causal
interpretations are not in general warranted. 

It is instructive to see the formal implications of the randomization for the properties of the least squares estimator, and to see how the randomization relates to the standard versions of the regression assumptions. 
To build this connection between the two repeated sampling paradigms it is very convenient to view the sample at hand as a random sample from an infinite population. This allows us to think of all the variables as random variables, with moments defined as population averages and with a distribution  induced by random sampling from this infinite population.
Define
\[ \tau=\mme\left[Y_i(1)-Y_i(0)\right],\aand \alpha=\mme\left[ Y_i(0)\right].\]
Then define the residual as
\[\varepsilon_i=Y_i(0)-\alpha+W_i\cdot \Bigl\{Y_i(1)-Y_i(0)-\tau\Bigr\}\]
\[\hskip2cm
=(1-W_i)\cdot \Bigl\{Y_i(0)-\mme[Y_i(0)]\Bigr\}+
W_i\cdot \Bigl\{Y_i(1)-\mme[Y_i(1)]\Bigr\}
.\]
This implies we can write the regression as in (\ref{regression}). 
Now the error term has a clear meaning as the difference between potential outcomes and their population expectation, rather than as the difference between the realized outcome and its conditional expectation given the treatment.
Moroever, the independence of $W_i$ and $(Y_i(0),Y_i(1))$, directly implied by the randomization, now has implications for the properties of the error term. Specifically, the randomization implies that the average residuals for treated and control units are zero:
\[ \mme[\varepsilon_i|W_i=0]=0,\ \ \ {\rm and}\  \mme[ \varepsilon_i|W_i=1]=0.\] 
Note that random assignment of the treatment does \underline{not} imply that the error term is independent of $W_i$. In fact, in general there will   be heteroskedasticity, and we need to use the Eicker-Huber-White robust standard errors to get valid confidence intervals.

It may appear that this is largely semantics, and that using regression methods  here makes no difference in practice. This is certainly true for estimation in this simple case without covariates, but not necessarily for inference. The conventional least squares approach suggests using the robust (Eicker-Huber-White) standard errors. Because the general robust variance estimator has no natural degrees-of-freedom adjustment, these standard robust variance estimators differs slightly from the Neyman unbiased variance estimator $\hat\mmv_\neyman$:
\begin{equation}
\hat{\mmv}_\robust=
\frac{\scs}{\nc}\cdot\frac{\nc-1}{\nc}+\frac{\sts}{\nt}\cdot\frac{\nt-1}{\nt}.
\label{vareen2}
\end{equation}
The Eicker-Huber-White variance estimator is not unbiased, and in settings where one of the treatment arms is rare, the difference may matter. 
For the Duflo-Hanna-Ryan data on the effect of teacher presence on educational achievement (Duflo, Hanna, and Ryan, 2012), this leads to
\begin{eqnarray*} \hat Y^\obs_i=&0.5805\ \  + & 0.2154\times W_i,\\
& (0.0256) &  (0.0308)\\
&& [0.0311]\end{eqnarray*}
with the Eicker-Huber-White standard errors in parentheses and the Neyman standard error in brackets.
Because both subsample sizes are large enough $(\nc=54$ and $\nt=53$), there is essentially no difference in the standard errors.
However, if we modify the sample so that there are $\nc=54$ control units but only $\nt=4$ treated units, the standard errors are quite different, 0.1215 for the Eicker-Huber-White standard errors, and 0.1400 for the Neyman standard errors.

Although there are refinements of the general Eicker-Huber-White variance estimator, there are none that are unbiased in general. The difference with the Neyman variance estimator relies on the fact that the only regressor in the Neyman variance estimator is a binary indicator.
Moreover, the Neyman variance estimator, fitting into the classic Behrens-Fisher problem, suggests using a t-distribution rather than a normal distribution with the degrees of freedom dependent on the size of the two treatment groups. See Imbens and Kolesar (2015) and Young (2016) for  recent discussions with illustrations how the distribution of the covariates matters for the standard errors.

\subsection{Regression Estimators with Additional Covariates}
\label{section:regression}

Now let us turn to the case with additional covariates beyond the treatment indicator $W_i$, with these additional covariates denoted by $X_i$. These additional covariates are not affected by the treatment by definition, that is, they are pre-treatment variables. Moreover, we assume here that these covariates did not affect the assignment, which we continue to assume is completely random. It is the presence of these covariates that often motivates using regression methods rather than simple differences by treatment status. 
There are generally three motivations for including these covariates into the analysis. First, they may improve the precision of the estimates. Second, they allow for estimation of average effects for subpopulations and in general for assessments of heterogeneity in treatment effects. Third, they may serve to remove biases in simple comparisons of means if the randomization was not adequate.
These are somewhat distinct, although related, goals, however, and regression methods are not necessarily the optimal choice for any of them.
In general, again, we wish to caution against the routine way in which regression methods are often applied here.

There are two ways  covariates are typically incorporated into the estimation strategy. First, they can be included additively through the regression model 
\begin{equation}\label{additive} Y^\obs_i=\alpha+\tau\cdot W_i+\beta'\dot X_i+\varepsilon_i.\end{equation}
Here $\dot X_i=X_i-\ox$ is the covariate measured in deviations from its mean. Using deviations from means does not affect the point estimates of $\tau$ or $\beta$, only that of the intercept $\alpha$, but this transformation of the covariates is convenient once we allow for interactions.
Estimating this regression function for the Duflo-Hanna-Ryan data changes the point estimate of the average effect to    $0.1921$  and leaves the standard error unchanged at   $0.0298$. The R-squared in the original regression was 
    0.3362, and
the two additional covariates increase this to 
    0.3596, which is not enough to make a difference in the standard error.

Second, we can allow for a model with a full set of interactions:
\begin{equation}\label{interactions} Y^\obs_i=\alpha+\tau\cdot W_i+\beta'\dot X_i+\gamma'\dot X_i\cdot W_i+\varepsilon_i.\end{equation}
In general the least squares estimates based on these regression functions are not unbiased
for the average treatment effects over the randomization distribution  given the finite population. There is one exception. If the covariates are all indicators and they partition the population, and we estimate the model with a full set of interactions, Equation (\ref{interactions}), then the least squares estimate of $\tau$ is unbiased for the average treatment effect. To see this, consider the simplest case with a single binary covariate. In that case we can think of average treatment effects $\tau_x$ for each value of $x$. We can also think of $\hat\tau_x$ estimated separately on the corresponding part of the subpopulation. If $\ox$ is the average value of $X_i$ in the sample, then
\[ \hat\tau=\hat\tau_1\cdot\ox+\hat\tau_0\cdot(1-\ox), \hskip1cm {\rm and}\ \ \hat\gamma=\hat \tau_1-\hat\tau_0.\]
Below in Section \ref{section:trees}, we discuss machine learning methods for partitioning the covariate space according to
treatment effect heterogeneity; if we construct indicators for the element of the partition derived according to an ``honest causal tree'' (Athey and Imbens, 2016) and incorporate them into (\ref{interactions}), then the resulting average treatment effect (estimated
on what Athey and Imbens (2016) refer to as the estimation sample) is unbiased
over the randomization distribution.  This result extends conceptually to the case where all regressors are indicators. In that case all least squares estimates are weighted averages of the within-cell estimated average effects.

If we are willing to make large sample approximations, we can also say something about the case with multivalued covariates. In that case, $\hat\tau$ is (asymptotically) unbiased for the average treatment effect. Moreoever, and this goes back to the first motivation for including covariates, the asymptotic variance for $\hat\tau$ is less than that of the simple difference estimator by a factor equal to  $1-R^2$ from including the covariates relative to not including the covariates. It is important that these two results do not rely on the regression model being true in the sense that the conditional expectation of $Y^\obs_i$ is actually linear in the covariates and the treatment indicator in the population. Because of the randomization there is zero correlation in the population between $W_i$ and the covariates $X_i$, which is sufficient for the lack of bias from including or excluding the covariates.
However, the large sample approximation needs to be taken seriously here. If in fact the covariates have very skewed distributions, the finite sample bias in the linear regression estimates may be substantial, as Freedman (2008) points out. At the same time, the gain in precision is often modest as the covariates often only have limited explanatory power.

The presence of non-zero values for $\gamma$ imply treatment effect heterogeneity. However, the interpretation of the $\gamma$ depends on the actual function form of the conditional expectations. Only if the covariates partition the population do these $\gamma$ have a clear interpretation as differences in average treatment effects. For that reason it may be easier to convert the covariates into indicator variables. It is unlikely that the goodness of fit of the regression model is much affected by such transformations, and both the interpretation and the finite sample unbiasedness would be improved by following that approach.

For the Duflo-Hanna-Ryan (2012) data, 
\begin{align*} \hat Y^\obs_i=&0.59+&0.192&\times W_i +&0.001 &\times X_{1i}-& 0.004&\times X_{2i}
-&0.006&\times X_{1i}\times W_i
\\
&(0.02)&(0.03)&&(0.002)&&(0.01)&&(0.003)&\\ 
+&0.017&\times X_{2i}\times W_i\\
&(0.01)\\
\end{align*}
The inclusion of the two covariates with the full set of interactions does not affect the point estimate of the average treatment effect, nor its standard error.

Alternatively, if we run the regression with an indicator for $X_{1i}>37$ (teacher score greater than the median), we get 
\begin{align*} \hat Y^\obs_i=&0.60+&0.188&\times W_i +&0.10 &\times {\bf 1}_{\{X_{1i}>37\}}&- 
0.06&\times {\bf 1}_{\{X_{1i}>37\}}\times W_i
\\
&(0.02)&(0.03)&&(0.05)&&(0.06)\\ \end{align*}
Now the coefficient on the interaction is directly interpretable as an estimate of the difference in the average effect for teachers with a score higher than 37 versus teachers with a score less than or equal to 37. Ultimately there is very little gain in precision in the estimator for the average treatment effect.

\section{The Analysis of  Stratified and Paired Randomized Experiments}
\label{section:stratification}

In this section we discuss the analyses for two generalizations of completely randomized experiments. First, consider stratified randomized experiments. In that case the covariate space is partitioned into a finite set. Within each of these subsets a completely randomized experiment is carried out.
In the extreme case where the partition is such that within each subset there are exactly two units, and the designs corresponds to randomly assigning exactly one of these two units to the treatment and the other to the control group we have a paired randomized experiment. 
Both these designs can be thought of as attempting to capture the gains from adjusting from observable differences between units by design, rather than by analysis as in the previous section. As such they capture the gains from ex post regression adjustment without the potential costs of linear regression, and therefore stratification is generally  preferable over regression adjustment. In the current section we discuss the analyses of such experiments, and in Section \ref{section:stratification} the design aspects.

\subsection{Stratified Randomized Experiments: Analysis}
\label{section:stratification_analysis}

In a stratified randomized experiment the covariate space is partitioned into a finite set of subsets. Within each of these subsets a completely randomized experiment is carried out, after which the results are combined. If we analyze the experiment using Neyman's repeated sampling approach the analysis of stratified randomized experiments is straightforward. Suppose there are $G$ strata within which we carry out a completely randomized experiment, possibly with varying treatment probabilities. Let $\tau_g$ be the average causal effect of the treatment for all units within stratum $g$. Within this stratum we can estimate the average effect as the difference in average outcomes for treated and control units:
\[ \hat\tau_g=\overline{Y}^\obs_{\ttt,g}-\overline{Y}^\obs_{\ccc,g},\]
and we can estimate the within-stratum variance, using the Neyman results, as
\[ \hat\mmv(\hat\tau_g)=\frac{s^2_{\ttt,g}}{N_{\ttt,g}}+\frac{s^2_{\ccc,j}}{N_{\ccc,g}},\]
where the $j$-subscript indexes the stratum. We can then estimate the overall average effect of the treatment by simply averaging the within-stratum estimates weighted by the stratum share $N_g/N$:
\[ \hat\tau=\sum_{g=1}^G \hat\tau_g\cdot \frac{N_g}{N},\hskip1cm {\rm with\ estimated\ variance\ }\ 
\hat\mmv_{\rm strat}(\hat\tau)=\sum_{g=1}^G  \hat\mmv(\hat\tau_g)\cdot\left( \frac{N_g}{N}\right)^2.\]

There is a special case that is of particular interest. Suppose the proportion of treated units is the same in all strata. In that case the estimator for the average treatment effect is equal
to the difference in means by treatment status,
\[ \hat\tau=\sum_{g=1}^G \hat\tau_g\cdot \frac{N_g}{N}=\oy^\obs_\ttt-\oy^\obs_\ccc,\]
 which is the estimator we used for the completely randomized experiment.
In general, however, the variance based on the completely randomized experiment set up,
\[ \hat\mmv_\neyman=\frac{s^2_{\ttt}}{N_{\ttt}}+\frac{s^2_{\ccc}}{N_{\ccc}},\]
will be conservative compared to the variance that takes into account the stratification, $\hat\mmv_{\rm strat}(\hat\tau)$: the latter takes into account the precision gain from stratification.

\subsection{Paired Randomized Experiments: Analysis}
\label{section:pairs}

Now let us consider a paired randomized experiment. Starting with $N$ units in our sample, $N/2$ pairs are constructed based on covariate values so that within the pairs the units are more similar in terms of covariate values. Then, within each pair a single unit is chosen at random to receive the active treatment and the other unit is assigned to the control group. The average treament effect within the pair is estimated as the difference in outcome for the treated unit and the control unit:
\[ \hat\tau_g=\sum_{i:G_{ig}=1,W_i=1} Y^\obs_i-\sum_{i:G_{ig}=1,W_i=0} Y^\obs_i.\]
The overall average effect is estimated as the average over the within-pair estimates:
\[ \hat\tau=\frac{1}{N/2}\sum_{g=1}^{N/2} \hat\tau_g=\oy^\obs_\ttt-\oy^\obs_\ccc.\]
So far this is similar to the analysis of a general stratified experiment, and conceptually the two designs are closely related.

The complications arise when estimating the variance of this estimator, as an estimator of the average effect over the strata, $\tau=\sum_{g=1}^{N/2} \tau_g\cdot 2/N$.
In the stratified randomized experiment case we estimated the variance in two steps, first estimating the within-stratum variance for stratum $g$ as
\[ \hat\mmv(\hat\tau_g)=\frac{s^2_{\ttt,g}}{N_{\ttt,g}}+\frac{s^2_{\ccc,g}}{N_{\ccc,g}},\]
followed by averaging this over the strata.
However, this variance estimator requires at least two treated and at least  two control units in each stratum, and thus is not feasible in the paired randomized experiment case with only one treated and one control unit in each stratum or pair.

Instead typically the following variance estimator is used:
\begin{equation}\label{var1} \hat\mmv(\hat\tau)=\frac{1}{N/2\cdot(N/2-1)}\sum_{g=1}^{N/2} \left(\hat\tau_g-\hat\tau\right)^2,\end{equation}
the variance of the $\hat\tau_g$ over the pairs. This variance estimator is conservative if viewed as an estimator of $\tau=\sum_{g=1}^{N/2} \tau_g\cdot 2/N$. However, suppose we view the pairs as being randomly drawn from a large super-population, with population average treatment effect equal to $\tau^*=\mme[\tau_g]$. Then the variance of $\hat\tau$, viewed as an estimator of $\tau^*$, can be estimated using $\hat\mmv(\hat\tau)$.

Because in this case the proportion of treated units is the same in each pair, namely $1/2$, we can also use the variance based on analysing this as a completely randomized experiment,
leading to:
\begin{equation}\label{var2} \hat\mmv_\neyman=\frac{s^2_{\ttt}}{N_{\ttt}}+\frac{s^2_{\ccc}}{N_{\ccc}}.\end{equation}
In general this will be conservative, and more so than necessary.

Let us illustrate this with data from the Children's Television Workshop experiment. See Imbens and Rubin (2015) for details. There are eight pairs of classrooms in this experiment, with one classroom in each pair shown the Electric Company, a children's television program.
The outcome is a post-test score, leading to
\[ \hat\tau=13.4,\hskip0.5cm (\widehat{\rm se}_{\rm pair}=4.6),\]
 where the standard error is calculated as  in Equation (\ref{var1}), taking into account the paired design. The variance estimate based on the interpretation as a completely randomized experiment as in Equation (\ref{var2}),  rather than a paired experiment, is $\hat{\rm se}_{\neyman}=7.8$, almost twice the size. There is a substantial gain from doing the paired randomized experiment in this case.

\section{The Design of Randomized Experiments and the Benefits of Stratification}
\label{section:stratification}

In this section we discuss some issue related to the design of randomized experiments. First we discuss the basic power calculations for completely randomized experiments. Second we discuss the benefits of stratification, and its limit, pairwise randomization, in terms of the expected precision of the resulting  experiments.  Finally we discuss issues related to re-randomization if one feels the randomization did not produce the desired balance in covariates between treatment and control groups. Ultimately our recommendation is that one should always stratify as much as possible, up to the point that each stratum contains at least two treated and two control units. 
Although there are in principle some benefits in terms of expected precision to using paired designs rather than stratified designs with two treated and two control units, these tend to be small and because  there are some real costs in terms of analyses we recommend the stratified rather than paired designs. 
If the stratification is done appropriately, there should be no need for re-randomization.

\subsection{Power Calculations}
\label{section:power}

In this section we look at some simple power calculations for randomized experiments. These are intended to be carried out prior to any experiment, in order to assess whether the proposed experiment has a reasonable chance of finding results of the size that one might reasonably expect. These analyses depend on a number of inputs, and can focus on various outputs. Here we largely focus on the formulation where the output is the minimum sample size required to find treatment effects of a pre-specified size with a pre-specified probability. Alternatively, one can also focus on the treatment size one would be likely to find given a particular sample size. For details on these and similar calculations a standard reference is Cohen (1988). See also
Murphy, Myors, and  Wollach, (2014).

Let us consider a simple case where for a sample of size $N$, we would observe
values for an outcome for the $N$ units, $Y^\obs_1,\ldots,Y^\obs_{N}$, and a treatment indicator $W_1,\ldots,W_N$. 
We are interested in testing the hypothesis that the average treatment effect is zero:
\[ H_0:\ \ \mathbb{E}[Y_i(1)-Y_i(0)]=0,\]
against the alternative that the average treatment effect differs from zero:
\[ H_a:\ \ \mathbb{E}[Y_i(1)-Y_i(0)]\neq 0.\]
We restrict the size of the test, the probability of rejecting the null hypothesis when it is in fact true, to be less than or equal to  $\alpha$. Often, following Fisher (1925), we choose $\alpha=0.05$ as the statistical significance level.  In addition,  we want the power of the test, the probability of rejecting the null when it is in fact false,  to be at least equal to $\beta$, in the case where the true average treatment effect is
$\tau=\mathbb{E}[Y_i(1)-Y_i(0)]$ for some prespecified value of $\tau$.
Let $\gamma=\sum_i W_i/N$ be the proportion of treated units. 
For simplicity we assume that the conditional outcome variance in each treatment arm is the same, $\sigma^2=\mmv(\yin)=\mmv(\yie)$.
We look for the minimum sample size $N=\nc+\nt$, as a function of
$\alpha$, $\beta$, $\tau$, $\sigma^2$, and $\gamma$.

To test the null hypothesis of no average treatment effect we look at the T-statistic 
\[ T=\frac{\overline{Y}^\obs_\ttt-\overline{Y}^\obs_\ccc}{\sqrt{S^2_Y/\nt+S^2_Y/\nc}}\approx \frac{\overline{Y}^\obs_\ttt-\overline{Y}^\obs_\ccc}{\sqrt{\sigma^2/\nt+\sigma^2/\nc}}.\]
We reject the null hypothesis of no difference if the absolute value of this t-statistic, $|T|$, exceeds $\Phi^{-1}\left(1-\alpha/2\right)$. Thus, if $\alpha=0.05$, the threshold would be $\Phi^{-1}\left(1-\alpha/2\right)=1.96$. We want the rejection probability to be at least $\beta$, given that the alternative hypothesis is true with the treatment effect equal to $\tau$. In general the difference in means minus the true treatment effect $\tau$, scaled by the standard error of that difference, has approximately a standard normal distribution:
\[\frac{\overline{Y}^\obs_\ttt-\overline{Y}^\obs_\ccc-\tau}{\sqrt{\sigma^2/\nt+\sigma^2/\nc}}\approx{\cal N}(0,1).\]
This implies that the t-statistic has an approximately normal distribution:
\[ T\approx{\cal N}\left(\frac{\tau}{\sqrt{\sigma^2/\nt+\sigma^2/\nc}},1\right).\]
Now,  a simple calculation implies that the null hypothesis will be rejected with probability 
\[ \pr\left(|T|>\Phi^{-1}\left(1-\alpha/2\right)\right)\approx
\Phi\left(-\Phi^{-1}\left(1-\alpha/2\right)+\frac{\tau}{\sqrt{\sigma^2/\nt+\sigma^2/\nc}}\right)\]
\[\hskip2cm +\Phi\left(-\Phi^{-1}\left(1-\alpha/2\right)-\frac{\tau}{\sqrt{\sigma^2/\nt+\sigma^2/\nc}}\right).\]
The second term is small, so we ignore it. 
Thus we want the   probability of the first term to be equal to $\beta$, 
which requires
\[ \beta=\Phi\left(-\Phi^{-1}\left(1-\alpha/2\right)+\frac{\tau}{\sqrt{\sigma^2/\nt+\sigma^2/\nc}}\right),\]
leading to
\[ \Phi^{-1}\left(\beta\right)=
-\Phi^{-1}\left(1-\alpha/2\right)+\frac{\tau\sqrt{N}\sqrt{(\gamma(1-\gamma)}}{\sigma}.\]
This leads to a required sample size 
\begin{equation}\label{een}
 N= \frac{ \left(\Phi^{-1}\left(\beta\right)+\Phi^{-1}\left(1-\alpha/2\right)\right)^2}
{(\tau^2/\sigma^2)\cdot\gamma\cdot(1-\gamma)}.
\end{equation}

For example, let us consider a setting close to the Lalonde data. The standard deviation of the outcome is approximately 6, although that may have been difficult to assess before the experiment. 
Suppose we choose $\gamma=0.5$ (equal sample sizes for treated and controls, which is optimal in the case with homoskedasticity, and typically close to optimal in other cases), $\alpha=0.05$ (test at 0.05 level). Suppose also that we are looking to be able to find an effect of 1 (thousand dollars), which is a substantial amount given the average pre-program earnings of these individuals, and that we choose $\beta=0.8$ (power of 0.8). Then
\[ N= \frac{ \left(\Phi^{-1}\left(\beta\right)+\Phi^{-1}\left(1-\alpha/2\right)\right)^2}
{(\tau^2/\sigma^2_Y)\cdot\gamma\cdot(1-\gamma)} = \frac{ \left(\Phi^{-1}\left(0.8\right)+\Phi^{-1}\left(0.975\right)\right)^2}
{0.167^2\cdot 0.5^2}=1,302,\]
so that the minimum sample size is 1,302, with 651 treated and 651 controls.
If the effect we wish to have power of 0.8 for is 2, then the required sample size would be substantially smaller, namely
282, split equally between 142 treated and 142 controls.

\subsection{Stratified Randomized Experiments: Benefits}
\label{section:stratification}

In this section we discuss the benefits of stratification in  randomized experiments.
Mostly this discussion is based on the special case where the ratio of the number of treated units to the total number of units is the same in each stratum. In this case the intended benefit of the stratification is to achieve balance in the covariates underlying the stratification. Suppose there are only two strata, containing, respectively, women and men. If the total sample size is 100, with 50 women and 50 men, and there are 60 individuals to be assigned to the treatment group and 40 to the control group, stratification would ensure that in the treatment group there are 30 women and 30 men, and the 20 of each sex in the control group. This would avoid a situation where, by chance, there were 25 women and 35 men in the treatment group, and 25 women and 15 men in the control group. If the outcomes were substantially correlated with the sex of the individual, such a random imbalance in the sex ratio in the two treatment groups would reduce the precision from the experiment.
Note that without stratification the experiment would still be valid, and, for example, still lead to exact p-values. Stratifying does not remove any bias, it simply leads more precise inferences than complete randomization.

Although it is well known that stratification on covariates is beneficial if based on covariates that are strongly correlated with the outcomes, there appears to be confusion in the literature concerning the benefits of stratification in small samples if this correlation is weak. 
Bruhn and McKenzie (2007) document this in a survey of researchers in development economics, but the confusion is also apparent in the statistics literature.
For example, Snedecor and Cochran (1989, page 101) write:
\begin{quote}
\textquotedblleft  If the criterion has no correlation with the response variable, a small loss in accuracy results from the pairing due to the adjustment for degrees of freedom. A substantial loss may even occur if the criterion is badly chosen so that member of a pair
 are negatively correlated.\textquotedblright
\end{quote}
Box, Hunter and Hunter (2005,  page 93) also suggest that there is a tradeoff in terms of accuracy or variance in the decision to stratify, writing:
\begin{quote}
\textquotedblleft Thus you would gain from the paired design only if the reduction in variance from pairing outweighed the effect of the decrease in the number of degrees of freedom of the $t$ distribution.\textquotedblright
\end{quote}
This is somewhat counterintuitive: if one stratifies on a covariate that is independent of all other variables, then stratification is obviously equivalent to complete randomization.
In the current section we  argue that this intuition is correct and that in fact there is no tradeoff. We present formal results that show that in terms of expected-squared-error, stratification (with the same treatment probabilities in each stratum) cannot be worse than complete randomization, even in small samples, and even with little, or even no, correlation between covariates and outcomes.  {Ex ante}, committing to stratification can only improve precision, not lower it. 
There is two important qualifications to this result.
First, {ex post}, given the joint distribution of the covariates in the sample, a particular stratification may be inferior to complete randomization. 
Second, the result requires that the sample can be viewed as a (stratified) random sample from an infinitely large population, with the expectation in the expected-squared-error taken over this population. This requirement guarantees that outcomes within strata cannot be negatively correlated.

The lack of any finite sample cost to {(ex ante)} stratification in terms of expected-squared-error contrasts with 
the potential cost of {ex post} stratification, or regression adjustment. {Ex post} adjustment for covariates 
through regression
may increase the finite sample variance, and in fact it will strictly increase the variance for any sample size, if the covariates have no predictive power at all.

However, there is a cost to stratifying on a variable that has no association with the potential outcomes.
Although there is no cost  to stratification in terms of the variance, there is a cost in terms of estimation of the variance. 
Because there are unbiased estimators for the variance, it follows that if the variance given stratification is less than or equal to the variance without stratification, it must be that the expectation of the estimated variance given stratification is less than or equal to the expectation of the estimated variance without stratification. However, the estimator for the variance given stratification typically has itself a larger variance, related to the
 degrees of freedom adjustment.
In our view this should not be interpreted, however, as an argument against stratification. 
One can always use the variance that ignores the stratification: this is conservative if the stratification did in fact reduce the variance.
See  Lynn and McCulloch (1992) for a similar argument in the context of paired randomized experiments.

We state the formal argument for a simplified case where we have a single binary covariate, 
$X_i\in\{\fff,\mmm\}$ (females and males).
we start with a large (infinitely large) superpopulation that expectations and variances refer to.
We will draw a sample of size
$N$ from this population and then assign treatments to each unit. 
For simplicity we assume that  $N/4$ is an integer. 
Each unit is characterized by a triple $(Y_i(0),Y_i(1),X_i)$, where $X_i$ is a binary indicator. In the superpopulation $X_i$ has a binomial distribution with support $\{\fff,\mmm\}$ (females and males) with ${\rm pr}(X_i=\fff)=1/2$.
Let $\mu_{\fff\ttt}=\mme[Y_i(1)|X_i=\fff]$, $\mu_{\fff\ccc}=\mme[Y_i(0)|X_i=\fff]$,
$\mu_{\mmm\ttt}=\mme[Y_i(1)|X_i=\mmm]$, and $\mu_{\mmm\ccc}=\mme[Y_i(0)|X_i=\mmm]$, and similarly for the variances.

We consider the following sampling scheme.   We randomly sample $N/2$ units from each of the two strata.
Given a sample of $X_1,\ldots, X_N$ we consider two randomization schemes.
In the first we randomly select $N/2$ units out of the sample of $N$ units to be assigned to the treatment group. We refer to this as the completely randomized assignment, $\mathbb{C}$.
Second, we consider the following stratified randomization scheme, denoted by $\mathbb{S}$. 
For the stratified design randomly select $N/4$ from each stratum to be assigned to the treatment, and assign the remainder to the control group. In both cases we estimate the average treatment effect as
\[ \hat\tau=\oy^\obs_\ttt-\oy^\obs_\ccc.\]

We consider the properties of this estimator over repeated randomizations, and repeated random samples from the population.
It follows  trivially that under both designs, the estimator is unbiased for the population average treatment effect under the randomization distribution.
The differences in performance between the estimators and the designs are solely the result of differences in the variances.
The exact variance for a completely randomized experiment can be written as
\[\mathbb{V}_\mathbb{C}
=
\frac{1}{4\cdot N}\cdot\left( (\mu_{\fff\ccc}-\mu_{\mmm\ccc})^2
+(\mu_{\fff\ttt}-\mu_{\mmm\ttt})^2\right)+
\frac{1}{N}\cdot
 \left(\sigma^2_{\fff\ttt}+\sigma^2_{\fff\ccc}\right)
+\frac{1}{N}\cdot
 \left(\sigma^2_{\mmm\ttt}+\sigma^2_{\mmm\ccc}\right).
\]
The variance for the corresponding stratified randomized experiment is
\[\mathbb{V}_\mathbb{S}=  \frac{1}{N}\cdot
 \left(\sigma^2_{\fff\ttt}+\sigma^2_{\fff\ccc}\right)
+\frac{1}{N}\cdot
 \left(\sigma^2_{\mmm\ttt}+\sigma^2_{\mmm\ccc}\right).
\]
Thus, the difference in the two variances is
\[\mathbb{V}_\mathbb{C}-\mathbb{V}_\mathbb{S}=
\frac{1}{4\cdot N}\cdot\left( (\mu_{\fff\ccc}-\mu_{\mmm\ccc})^2
+(\mu_{\fff\ttt}-\mu_{\mmm\ttt})^2\right)
\geq 0
.\]
Therefore, stratification leads to variances that cannot be higher than those under
a completely randomized experiment. There can only be equality if neither of the potential outcomes is correlated with the covariate, and
$\mu_{\fff\ccc}=\mu_{\mmm\ccc}$ and $\mu_{\fff\ttt}=\mu_{\mmm\ttt}$. This is the main argument for our recommendation that one should always stratify.

The inability to rank the conditional variance is useful in understanding the Snedecor and Cochran quote in the introduction. If the strata are defined in terms of a continuous covariate, than in a particular sample, it is possible that stratification leads to larger variances conditional on the covariate values (and in the special case of paired experiments, to negative correlations within pairs). That is not possible on average, that is, over repeated samples randomly drawn from large strata, rather than conditional on the covariate values in a single sample. As mentioned before, the large strata qualification here is important: if the strata we draw from are small, say litters of puppies, it may well be that the within-stratum correlation is negative, but that is not possible if all the strata are large: in that case the correlation has to be non-negative.

Now let us
consider two estimators for the variance. First 
define, for $w=\ccc,\ttt$, and $x={\rm f},{\rm m}$,
\[s_{xw}^2=\frac{1}{N_{xw}-1}\sum_{i:W_i=1_{\{w=\ttt\}},X_i=x} \left(Y^\obs_i-\oy^\obs_{xw}\right)^2
\hskip0.3cm {\rm and}\ \ 
s_w^2=\frac{1}{N_w-1}\sum_{i:W_i=1_{\{w=\ttt\}}} \left(Y^\obs_i-\oy^\obs_w\right)^2.
\]
The natural estimator for the variance under the completely randomized experiment is:
\[ \hmmv_\mathbb{C}=\frac{\scs}{\nc}+\frac{\sts}{\nt}, \hskip1cm {\rm with}\ \ \mme[\hmmv_\mathbb{C}]=\mmv_\mmc.\]
For a stratified randomized experiment the natural variance estimator, taking into account the stratification,
is:
\[ \hmmv_\mathbb{S}=\frac{\nf}{\nf+\nm}\cdot\left(\frac{s_{\fff\ccc}^2}{\nfn}+\frac{s_{\fff\ttt}^2}{\nfe}\right)
+
\frac{\nm}{\nf+\nm}\cdot\left(\frac{s_{\mmm\ccc}^2}{\nmn}+\frac{s_{\mmm\ttt}^2}{\nme}\right)
\hskip1cm {\rm 
 with}\ \ 
 \mme[\hmmv_\mms]=\mmv_\mms.\]
Hence, $ \mme[\hmmv_\mms]\leq  \mme[\hmmv_\mmc]$. Nevertheless, in a particular  sample,
with values $(\by,\bw,\bx)$, it may well be the case that
the realized value of the  completely randomized variance estimator 
$\hmmv_\mmc(\by,\bw,\bx)$ is less  than that of the stratified variance  $\hmmv_\mms(\by,\bw,\bx)$. To be more specific, consider the case where the stratification is not related to the potential outcomes at all. In that case the two variances are identical in expectation, $\mme[\hat\mmv_\mathbb{S}]=\mme[\hat\mmv_\mathbb{C}]$, but the variance of $\hmmv_\mms$ is larger than the variance of $\hmmv_\mmc$,
$\mmv(\hat\mmv_\mathbb{S})<\mmv(\hat\mmv_\mathbb{C})$. As a result the power of a t-test based on $\hmmv_\mms$ will be slightly lower than the power of a t-test based on $\hmmv_\mmc$. Nevertheless, in practice we recommend to always stratify whenever possible.

\subsection{Re-randomization}
\label{section:re-randomization}

Suppose one is conducting a randomized experiment. For the study population the researcher has collected some background characteristics and has decided to assign the units to the treatment or control group completely at random. Although this would in general not be optimal, it may be that the researcher decided it was not worth the effort investigating a better design, and just went ahead with the complete randomization. Now, however, suppose that after the random assigmnent has been decided, but prior to the actual implementation of the assignment, the researcher compares average pretreatment values by treatment status. In expectation these should be identical for all covariates, but obviously in reality these will differ somewhat. Now suppose that one of the most important covariates does actually show a substantial difference between the assigned treatment and control group. It need not be, although it may be statistically significant at conventional levels even if the randomization was done properly, simply because there is a substantial number of covariates, or simply by chance. What should one do in that case? More specifically, should one go back to the drawing board and re-randomize the treatment assignment so that the important covariate is better balanced? This question of re-randomization has received some attention in the empirical development literature. One paper that raised the question forcefully in this literature is Bruhn and McKenzie (2007).
Theoretical paper discussing some of the formal aspects are Morgan and Rubin  (2012) and Banerjee, Snowberg, and Chassang (2016).

Here we offer some comments. First of all, implicitly many designs for randomized experiments can be thought of as based on re-randomization. Consider the case where the population of $N=100$ individuals consists of 50 women and 50 men. Suppose we do a completely randomized experiment, with 60 individuals to be assigned to the treatment group and the remaining 40 assigned to the control group. Now suppose we reject and re-randomize any randomization vector that does not correspond to 30 men and 30 women being assigned to the treatment group. Then, in an indirect manner, we end up with a stratified randomized experiment that we know how to analyze, and that in general offers better sampling properties in terms of variance. The point is that in this case the re-randomization does not create any complications, although the appropriate analysis given the re-randomization is different from the one based on ignoring the re-randomization. Specifically, p-values need to be adjusted for the re-randomization, although ignoring the adjustment simply leads to conservative p-values. Both statements hold more generally.

In order for the subsequent analysis to be able to take account of the re-randomization, however, the details of the re-randomization need to be spelled out. This is most easily seen if we consider a Fisher exact p-value analysis. In order to calculate the exact p-value we need to know the exact distribution of the assignment vector. In the case of possible re-randomization we would therefore need to know exactly which assignment vectors would be subject to re-randomization, and which would be viewed as acceptable. The actual criterion may be complicated, and involve calculation of t-statistics for differences in average covariates between treatment groups, but it needs to be completely spelled out in order for the exact p-value calculation to be feasible.
Doing this is ultimately equivalent to designing an experiment that guarantees more balance, and it would most likely take  a form close to that of a stratified randomized experiment. We recommend simply taking care in the original design so that assignments that correspond to unacceptable balance are ruled out from the outset, rather than ruled out ex post which complicates inference.

\section{The Analysis of Clustered Randomized Experiments}
\label{section:clustering}

In this section we discuss clustered randomized experiments. Instead of assigning treatments at the unit level, in this setting the population is first partitioned into a number of clusters. Then all units in a cluster are then assigned  to the same treatment level. 
Clusters may take the form of schools, where within a school district a number of schools are randomly assigned to an educational intervention rather than individual students, or villages, or states, or other geographical entities.
For general discussions, see Donner (1987), Gail, Mark, Carroll, Green, and Pee (1996), and Murray (2012).

Given a fixed sample size, this design is in general not as efficient as a completely randomized design or a stratified randomized design. The motivation for such clustered designs is  different. One motivation is that in some cases there may be interference between units at the unit-level. If there is no interference between units in different clusters, then the cluster-level randomization may allow for simple, no-interference type analyses, whereas a unit-level analysis would require accounting for the within-cluster interactions. A second motivation is that in many cases it is easier to sample units at the cluster level. For the same cost, or level of effort, it may be therefore be possible to collect data on a larger number of units.

In practice there are quite different settings where clustered randomization may take place. In some cases the number of units per cluster is similar, for example in educational settings where the clusters are classrooms. In other settings where clusters are geographical units, e.g., states, or towns, there may be a substantive amount of variation in cluster size. Although theoretically this does not make much of a difference, in practice it can affect what effective strategies are available for dealing with the clustering.
In the first case our main recommendation is to include analyses that are based on the cluster as the unit of analysis. Although more sophisticated analyses may be more informative than simple analyses using the clusters as units, it is rare that these differences in precision are substantial, and a cluster-based analysis has the virtue of great transparency.
 Analyzing the data at the unit-level has the benefit that one can directly take into account unit-level characteristics. In practice, however, including unit-level characteristics generally improves precision  by a relatively modest amount compared to including cluster-averages as covariates in a cluster-level analysis, so our recommendation is to focus primarily on cluster-level analyses. For the second case where there is substantial variation in cluster sizes, a key component of our recommended strategy is to focus  on analyses with cluster-averages as the target in addition to analysis with unit-averages that may be the main target of interest. The former may be much easier to estimate in settings with a substantial amount of heterogeneity in cluster sizes.

\subsection{The Choice of Estimand in Clustered Randomized Experiments}
\label{section:estimands}

As before, let $G$ be the number of clusters, and let $G_{ig}\in\{0,1\}$, $i=1,\ldots,N$, $g=1,\ldots,G$ denote the binary indicator that unit $i$ belongs to cluster $g$.  $N_g=\sum_{i=1}^N G_{ig}$ is the number of units in cluster $g$, so that $N_g/N$ is the share of cluster $g$ in the sample. $W_i$ continues to denote the treatment assignment for unit $i$, but now $\ow_g$ denotes the average value of the treatment assignment for all units in cluster $g$, so that by the definition of clustered randomized assignment, $\ow_g\in\{0,1\}$. Let $G_{\ttt}$ be the number of treated clusters, and $G_{\ccc}=G-G_\ttt$ the number of control clusters.

The first issue in clustered randomized experiments is that there may be different estimands to consider. 
One natural estimand is the overall population average treatment effect,
\[ \tau^\pop=\frac{1}{N}\sum_{i=1}^N\Bigl(\yie-\yin\Bigr),\]
where we average over all units in the population.
A second estimand is the unweighted average of the within-cluster average effects:
\[ \tau^\mmc=\frac{1}{G}\sum_{g=1}^G \tau_g,
\hskip1cm {\rm where}\ \ 
\tau_g=\frac{1}{N_g}\sum_{i:C_{ig}=1}\Bigl(\yie-\yin\Bigr).\]
We can think of $\tau^\mmc$ as a weighted average of the unit-level treatment effects, with the weight for units in cluster $g$ proportional to the inverse of the cluster sample size, $1/N_g$. Similarly, the population average treatment effect can be thought of as a weighted average of the cluster-level average treatment effects with weights proportional to the cluster sample size $N_g$.

There are two issues regarding the choice of estimand. One is which of the estimands is of most substantive interest. In many cases this will be the unweighted, population average treatment effect $\tau^\pop$. A second issue is the ease and informativeness of any analysis. Because the randomization is at the cluster level, simply aggregating unit-level outcomes to cluster-level averages simplifies the analyis substantially: all the methods developed for completely randomized experiments apply directly to a cluster-level analysis for clustered randomized experiments. In addition, inferences for $\tau^\mmc$ are often much more precise than inferences for $\tau^\pop$ in cases where there are a few large clusters and many small clusters. Consider an extreme case where there is one extremely large cluster, that in terms of size is larger then the other clusters combined. Inference for $\tau^\pop$ in that case is difficult because all the units in this mega-cluster will always be in the same treatment group. Inference for $\tau^\mmc$, on the other hand, may well be precise.
Moreoever, if the substantive question is one of testing for the presence of any treatment effect, answering this question by focusing on a statistic that averages over clusters without weighting is just as valid as comparing weighted averages over clusters.

In practice a researcher may therefore want to report analyses for $\tau^\pop$ in combination with analyses for $\tau^\mmc$. In cases where $\tau^\pop$ is the estimand that is of most substantive interest, the more precise inferences for $\tau^\mmc$ may complement the noisy analyses for 
the substantively more interesting
$\tau^\pop$.

\subsection{Point Estimation in Clustered Randomized Experiments}
\label{section:clustering_analysis}

Now let us consider the analysis of cluster randomized experiments. We focus on the case where unit-level outcomes and possibly covariates are available.  
The first choice facing the researcher concerns the choice of the unit of analysis. One can analyze the data at the unit level or at the cluster level. We first do the latter, and then return to the former.

If we are interested in the average effect $\tau^\mmc$, we can directly use the methods for completely randomized experiments discussed in Section \ref{section:randomizedexperiments}. 
Let $\oy_g^\obs$ be the average of the observed outcomes in cluster $g$. We can simply average the averages for the treated and control clusters:
\[ \hat\tau^\mmc=\frac{1}{G_\ttt}\sum_{g:\ow_g=1} \oy^\obs_g-
 \frac{1}{G_\ccc}\sum_{g:\ow_g=0} \oy^\obs_g.\]
The variance of this estimator can be estimated as
\[ \hat\mmv(\hat\tau^\mmc)=\frac{s^2_{\mmc,\ccc}}{G_\ccc}+
\frac{s^2_{\mmc,\ttt}}{G_\ttt},\]
where the variance for the averages of 
\[ s^2_{\mmc,\ccc}=\frac{1}{G_\ccc-1}\sum_{g:\ow_g=0} \left(\oy^\obs_g-\frac{1}{G_\ttt}\sum_{g':\ow_{g'}=1} \oy^\obs_{g'}\right)^2,\]
and similarly for $s^2_{\mmc,\ccc}$.
We can also get the same estimates using regression methods for the regression function
\begin{equation}\label{regress2} \oy^\obs_g=\alpha+\tau^\mmc\cdot \ow_g+\eta_g.\end{equation}
We can generalize the specification of this regression function to include cluster-level covariates, including cluster characteristics or averages of unit-level characteristics.

Using a unit-level analysis obtaining an estimate for $\tau^\mmc$ is more complicated.
Consider the regression
\begin{equation}\label{regress} Y^\obs_i=\alpha+\tau\cdot W_i+\varepsilon_i.\end{equation}
We can estimate this regression function using weighted least squares with the weight for unit $i$, belonging to cluster $g(i)$, equal to $1/N_{g(i)}$ as in the Cox (1956) analysis of weighted randomized experiments. This weighted least squares estimator is identical to $\hat\tau^\mmc$.

Now consider the case where we are interested in $\tau^\pop$. In that case we can estimate the regression function in (\ref{regress}) without any weights. Alternatively, we can get the same numerical answer by
estimating the regression (\ref{regress2})
at the cluster level with weights proportional to the cluster sample sizes $N_g$. To get the variance for the estimator for the population average  treatment effect we can use the unit-level regression, but we need to take into account the clustering. We can do so using the robust clustering standard errors proposed by Liang and Zeger (1986).
Let $\hat\alpha$ and $\hat\tau$ be the least squares estimators for $\alpha$ and $\tau$ based on (\ref{regress}), and let $\hat\varepsilon_i=Y^\obs_i-\hat\alpha-\hat\tau\cdot W_i$ be the residual. Then 
the covariance matrix for $(\hat\alpha,\hat\tau)$ can be estimated
 as
\[ \left(\sum_{i=1}^N
\left(\begin{array}{cc} 1 & W_i\\ W_i &W_i\end{array}\right)\right)
^{-1}
\left(\sum_{g=1}^G
\sum_{i:C_{ig}=1}\left(\begin{array}{c} \hat\varepsilon_i
\\ W_i\cdot \hat\varepsilon_i\end{array}\right)
\sum_{i:C_{ig}=1}\left(\begin{array}{c} \hat\varepsilon_i
\\ W_i\cdot \hat\varepsilon_i\end{array}\right)'
\right)
\left(\sum_{i=1}^N
\left(\begin{array}{cc} 1 & W_i\\ W_i &W_i\end{array}\right)
\right)^{-1}.\]
The key difference with the Eicker-Huber-White robust standard errors is that before taking the outer product of the product of the residuals and the covariates they are summed up within clusters.
This cluster-robust variance estimator is implemented in many regression software packages, sometimes with {\it ad hoc} degrees of freedom adjustments.

If we compare unit-level and cluster-level analyses in the form described so far, 
our preference is for cluster-level analysis, as it is more transparent and more
directly linked to the randomization framework.  However, unit-level analysis allows the analyst to impose additional modeling
assumptions; for example, a unit-level regression can incorporate covariates and impose additional assumptions, such as
restricting the effect of covariates to be common across clusters.  If justified, imposing such restrictions can increase efficiency.
One could accomplish the same goal by first doing a unit-level regression and constructing residuals for each unit, and then performing cluster-level analysis on the residuals, but at that point the inference would become more complex and depart from the pure
randomization-based analysis, reducing the benefits of a cluster-based approach.

\subsection{Clustered Sampling and Completely Randomized Experiments}

A second issue related to clustering is that the original sample may have been obtained through clustered sampling. This issue is discussed in more detail in Abadie, Athey, Imbens and Wooldridge (2016). Suppose we have a large population. The population is divided into $G$ clusters, as in the previous discussion. Instead of a random sample from this population,  we first sample a number of clusters from the population of clusters. Within each of the sampled clusters we sample a fixed fraction of the units within that cluster. Given our sample we conduct a completely randomized experiment, without regard to the  cluster these units belong to.

There is a subtle issue involved in defining what the estimand is. The first alternative is to focus on the sample average treatment effect, that is, the average difference for the two potential outcomes over all the units in the sample.  A second alternative is to analyze the population average treatment effect for all the units in the population, including those in non-sampled clusters. For both alternatives, the simple difference in average outcomes by treatment status is unbiased for the estimand.

Abadie, Athey, Imbens and Wooldridge (2016) show that we are interested in the sample average treatment effect, we can ignore the clustering and use the conventional Neyman variance estimator discussed in Section \ref{section:randomizedexperiments}. In contrast, if we are interested in the population average treatment effect, we need to take into account the implications of the clustering sampling design. We can adjust the standard errors for the clustered sampling by using the Liang-Zeger (1986) clustered standard errors.

\section{Noncompliance in Randomized Experiments}
\label{section:noncompliance}

Even if a randomized experiment is well designed, there may be complications in the implementation. One of the most common of these complications is {non-compliance}. Some units assigned to the treatment group may end up not taking the treatment, and some units assigned to the treatment group may manage to acquire the active treatment. If there are only violations of the treatment assignment of the first type, we refer to it as {one-sided non-compliance}. This may arise when individuals assigned to the control groups can be effectively be embargoed from the active treatment. If some units assigned to the control group do manage to receive the active treatment we have { two-sided non-compliance}. 

The concern is that noncompliance is not random or accidental, but the result of systematic differences in behavior or characteristics between units. Units who are assigned to the treatment but who choose  not receive it may do so because they are different from the units assigned to the treatment who do receive it. These differences may be associated with the outcomes of interest, thereby invalidating simple comparisons of outcomes by treatment received. In other words, the randomization that validates comparisons by treatment status does not validate comparisons by post-treatment variables such as the treatment received. These issues come up both in randomized experiments as well as in observational studies. The general term for these complications in the econometric literature is {endogeneity} of the receipt of treatment. Random assignment ensures that the assignment to treatment is exogenous, but it does not bear on the exogeneity of the receipt of treatment if the receipt of treatment is different from the assignment to treatment.

In this chapter, we discuss three distinct approaches to dealing with non-compliance, all of which are valid under fairly weak assumptions. First, one can ignore the actual receipt of the treatment and focus on the causal effects of assignment to the treatment, in an { intention-to-treat} analysis. Second, we can use instrumental variables methods to estimate the {local average treatment effect}, the causal effect of the receipt of treatment for the subpopulation of {compliers}. Third, we can use a {partial identification} or {bounds} analysis to obtain the range of values for the average causal effect of the receipt of treatment for the full population. 
Another approach, not further discussed here, is the randomization-based approach to instrumental variables developed in Imbens and Rosenbaum (2005).
There are also two types of analyses that require much stronger assumptions in order to be valid. The first of these is an {as-treated} analysis, where units are compared by the treatment received; this relies on an unconfoundedness or selection-on-observables assumption. A second type of analysis is a {per protocol} analysis, where units are dropped who do not receive the treatment they were assigned to. 

We need some additional notation in this section. Let $Z_i\in\{0,1\}$ denote the randomly assigned treatment. We generalize the notation for the treatment received, to reflect its status as an (intermediate) outcome. Let $W_i(z)\in\{0,1\}$ denote the potential treatment outcome given assignment $z$, with $W^\obs_i=W_i(Z_i)$ the realized value for the treatment received. For the outcome of primary interest, there are different set ups possible. 
One approach, e.g., Angrist, Imbens and Rubin (1996), is to let $Y_i(z,w)$ denote the potential outcome  corresponding to  assignment $z$ and treatment received $w$. Alternatively, we could index the potential outcomes solely by the assignment, with $\tilde Y_i(z)$ denoting the outcome corresponding to the treatment assigned to unit $i$. The two notations are closely related, with $\tilde Y_i(z)=Y_i(z,W_i(z))$. Here we mainly use the first set up. The realized outcome is $Y^\obs_i=Y_i(Z_i,W_i(Z_i))=\tilde Y_i(Z_i)$.
To simplify notation, we index sample sizes, averages, and variances by $0,1$ when they are indexed by values of the assignment $Z_i$, and by $\tc,\ttt$ when they are indexed by values of the treatment received $W_i$. For example, $\oy_{0,\ttt}^\obs$ is the average of the observed outcome for units assigned to the control group ($Z_i=0$) but who received the active treatment ($W^\obs_i=1$).

\subsection{Intention-To-Treat Analyses}
\label{section:itt}

In an intention-to-treat analysis the receipt of treatment is ignored, and outcomes are compared by the assignment to treatment (Imbens and Rubin, 2015; Fisher et al, 2000). The intention-to-treat effect is the average effect of the assignment to treatment. In terms of the notation introduced above, the estimand is
\[\tau^\itt=\frac{1}{N}\sum_{i=1}^N \Bigl( Y_i(1,W_i(1))-Y_i(0,W_i(0))\Bigr).\]
We can estimate this using the difference in averages of realized outcomes by treatment assignment:
\[\hat\tau^\itt=\oy^\obs_1-\oy^\obs_0,\hskip1cm {\rm 
where}
\ \ \oy^\obs_z=\frac{1}{N_z}\sum_{i:Z_i=z} Y^\obs_i\ \ {\rm for}\  z=0,1.\]
To construct valid confidence intervals for $\tau^\itt$ we can use the standard methods discussed in Section ref{section:neyman}. The exact variance for $\hat\tau^\itt$ is
\[ \mmv\left(\hat\tau^\itt\right)=
\frac{S^2_0}{N_0}+\frac{S^2_1}{N_1}-\frac{S_{01}^2}{N},
\]
where $S^2_0$ and $S^2_1$ are the variances of $Y_i(0,W_i(0))$ and $Y_i(1,W_i(1))$ in the sample,
defined as:
\[S_0^2=\frac{1}{N-1}\sum_{i=1}^{N}\Bigl(Y_i(0,W_i(0))-\oy(0)\Bigr)^2,\aand 
S_1^2=\frac{1}{N-1}\sum_{i=1}^{N}\Bigl(Y_i(1,W_i(1))-\oy(1)\Bigr)^2,
\]
and $S_{01}^2$ is the sample variance of the unit-level
treatment effects, defined as:
\[S_{01}^2=\frac{1}{N-1}\sum_{i=1}^{N}
\Bigl(Y_i(1,W_i(1))-Y_i(0,W_i(0))-
(\oy(1)-\oy(0))\Bigr)^2.\]
We can estimate the first two terms as
\[ s^2_0
=\frac{1}{N_0-1}\sum_{i:Z_i=0} \left(Y_i(0,W_i(0))
-\oy^\obs_0\right)^2 ,\]
and
\[ s^2_1=\frac{1}{N_1-1}\sum_{i:Z_i=1} \left(Y_i^\obs
-\oy_1\right)^2.\]
As discussed in Section \ref{section:neyman}, the third term, $S_{01}^2$
 is generally impossible
to estimate consistently because we  never observe both $Y_i(1,W_i(1))$ and $Y_i(0,W_i(0))$
for the same unit. 
In practice we therefore use the estimator for $\mmv\left(\hat\tau^\itt\right)$ based on estimating the first two terms by $s^2_0$ and $s^2_1$, and ignoring the third term,
\[
\hat{\mmv}(\hat\tau^\itt)=
\frac{s^2_0}{N_0}+\frac{s^2_1}{N_1}.
\]
This leads to valid confidence intervals in large samples, justified by the randomization and sutva without additional assumptions. 

The main drawback associated with the intention-to-treat approach is that the corresponding estimand  is typically not the object of primary interest. The researcher may be interested in settings where the assignment mechanism may be different, and the incentives for individuals to take the treatment might change. 
For example, in medical drug trials the compliance rate is often very different from what would happen if a drug is released to the general population. In the trial phase individuals, knowing that the efficacy of the drug has not been established, may be more likely to stop adhering to the protocol.
 As a result  the intention-to-treat effect would not provide much guidance to the effects in the new setting. 
In other words, intention-to-treat effects may have poor external validity. The presumption is that causal effects of the receipt of treatment are more generalizable to other settings, though of course there is no formal result that proves that this is so.

\subsection{Local Average Treatment Effects}
\label{section:late}

An alternative approach that deals directly with the non-compliance is to use instrumental variables methods and related methods based on principal stratification (Frangakis and Rubin, 2002; Barnard, Du, Hill, and Rubin, 1998). Bloom (1984), Zelen (1979, 1990), Baker  and Lindeman (1994), Cuzick,  Edwards,  and Segnan (1997), contain early and independent discussions of the instrumental variables approach, some  in the special case of one-sided non-compliance, and Imbens and Angrist (1994), Angrist, Imbens and Rubin (1996) develop the general  set up in the potential outcomes framework.
 See also Imbens and Rubin (2015) and Lui (2011) for  textbook discussions and 
Baker, Kramer, and Lindeman (2016) for a biostatistical perspective. The first step is to consider the possible patterns of compliance behavior. Let $C_i\in\{c,d,a,n\}$ denote the compliance behavior, where
\[ C_i=\left\{
\begin{array}{ll}
c\hskip1cm & {\rm if}\ W_i(0)=0,W_i(1)=1,\\
d\hskip1cm & {\rm if}\ W_i(0)=1,W_i(1)=0,\\
a\hskip1cm & {\rm if}\ W_i(0)=1,W_i(1)=1,\\
n\hskip1cm & {\rm if}\ W_i(0)=0,W_i(1)=0,\\
\end{array}\right.
\] 
where $c$ stands for complier, $d$ for defier, $n$ for never-taker, and $a$ for always-taker.
These labels are just definitional, not requiring any assumptions.

Now we consider two key assumptions. The first is { monotonicity} (Imbens and Angrist, 1994), or no-defiance, which requires
\[ W_i(1)\geq W_i(0).\]
This rules out the presence of defiers, units who always (that is, whether assignmed to control or treatment), do the opposite of their assignment. In the setting we consider in this chapter, where the instrument is the random assignment to treatment, this appears a very plausible assumption: assigning someone to the active treatment increases the incentive to take the active treatment, and it would appear unusal for there to be many units who would respond to this increase in incentives by declining to take the treatment where they would otherwise have done so. In other settings monotonicity may be a more controversial assumption. For example, in studies in criminal justice researchers have used random assignment of cases to judges to identify the causal effect of prison terms on recidivism (reference). In that case even if one judge is more strict than another in the sense that the first judge has a higher rate of sentencing individuals to prison terms, it is not necessarily the case that any individual who would be sentenced to time in prison by the on-average more lenient judge would also be sentenced to prison by the stricter judge.

The second key assumption is generally referred to as the { exclusion restriction}. It requires that there is no direct effect of the assignment on the outcome without passing through the receipt of treatment. Formally, using the form used in Angrist, Imbens and Rubin (1996),
\[ Y_i(z,w)=Y_i(z',w), \ \ {\rm for\ all\ } z,z',w.\]
The key components on the assumption is that for never-takers,
\[ Y_i(0,0)=Y_i(1,0),\hskip1cm {\rm  and\ for\ always-takers\ }Y_i(0,1)=Y_i(1,1).\] For compliers and defiers the assumption is essentially about the interpretation of the causal effect of the assignment to treatment to the causal effect of the receipt of treatment. The exclusion restriction is a strong one, and its plausibility needs to be argued on a case-by-case basis. It is not justified by, and in fact not related to, the random assignment. Given the exclusion restriction we can drop the dependence of the potential outcomes on $z$, and simply write $Y_i(w)$, for $w=0,1$.

Given the monotonicity assumption and the exclusion restriction we can identify the average causal effect of the receipt of treatment on the outcome, what is known as the {local average treatment effect} (Imbens and Angrist, 1994):
\[\tau^{\rm late}=\mme[Y_i(1)-Y_i(0)|C_i=c]=\frac{\mme[Y_i^\obs|Z_i=1]-\mme[Y_i^\obs|Z_i=0]}{\mme[W_i^\obs|Z_i=1]-\mme[W_i^\obs|Z_i=0]}.\]
Given the setting it is clear that we cannot identify the average effect for always-takers or neve-takers without additional assumptions: we do not observe outcomes for always-takers without the receipt of treatment, and we do not observe outcomes for never-takers given receipt of treatment. As a result we need assumptions to extrapolate the treatment effects for compliers to other compliance groups in order to identify the overall average treatment effect.

\subsection{Generalizing the Local Average Treatment Effect}
\label{section:generallate}

One major concern with the local average treatment effect is that it reflects only on a subpopulation, the compliers. In many cases the researcher may be more interested in the overall average effect of the treatment. Here we discuss some supplementary analyses that can be done to assess the generalizability of the local average treatment effect. This section builds on the discussions in Angrist (2004),
Hirano, Imbens, Rubin, and Zhou (2000), Imbens and Rubin (1997ab)
 and Bertanha and Imbens (2014). The Bertanha and Imbens (2015) discussion is primarily in the context of fuzzy regression discontinuity designs, but their results apply directly to other instrumental variables settings.

We use the same set up as in the previous section, but explicitly allow for the presence of exogenous covariates $X_i$. Instead of using instrumental variables methods to estimate the local average treatment effect  an alternative approach is to adjust for differences in the covariate to estimate the average effect of the treatment, assuming unconfoundedness:
\[ W_i\ \indep\ \Bigl(Y_i(0),Y_i(1)\Bigr)\ \Bigl|\ X_i.\]
If this assumption is valid, we can estimate the average effect of the treatment, as well as average effects for any subpopulation using an as-treated analysis.
One natural analysis is to compare the local average treatment effect to the covariate-adjusted difference by  treatment status. A formal comparison of the two estimates, in a linear model setting, would be a Hausman test (Hausman, 1983). In the absence of covariates the Hausman test would be testing the equality
\[ 
\frac{\pi_\ta}{\pi_\ta+\pi_\tc\cdot p_z}\cdot
 \Bigl(\mme[Y_i(1)|\ggc_i=\ta]
-\mme[Y_i(1)|\ggc_i=\tc]
\Bigr)\]
\[\hskip2cm =
\frac{\pi_\tn}{\pi_\tn+\pi_\tc\cdot (1-p_z)}\cdot
 \Bigl(\mme[Y_i(0)|\ggc_i=\tn]
-\mme[Y_i(0)|\ggc_i=\tc]
\Bigr),\]
where $\pi_\ta$, $\pi_\tc$, and $\pi_\tn$ are the population shares of always-takers, compliers, and never-takers respectively. This equality is difficult to interpret. A particular weighted average of the difference between the expected outcomes given treatment for always-takers and compliers is equal to a weighted average of the difference between the expected outcomes without treatment for compliers and never-takers.

Compared to the Hausman test a more natural and interpretable approach is to test the equality of the unweighted differences, between always-takers and treated compliers and never-takers and not-treated compliers,
\[ \mme[Y_i(1)|G_i=\ta]-\mme[Y_i(1)|G_i=\tc]=
\mme[Y_i(0)|G_i=\tn]-\mme[Y_i(0)|G_i=\tc],\]
as suggested in Angrist (2004).

Bertanha and Imbens suggest testing the pair of equalities, rather than just the difference,
\[ \mme[Y_i(1)|G_i=\ta]-\mme[Y_i(1)|G_i=\tc]=0,\hskip1cm {\rm and}\ \ 
\mme[Y_i(0)|G_i=\tn]-\mme[Y_i(0)|G_i=\tc]=0.\]
If this pair of equalities hold, possibly after adjusting for differences in the covariates, it means that always-takers are comparable to compliers given the treatment, and never-takers are comparable to compliers without the treatment. That would suggest that always-takers without the treatment might also be comparable to compliers without the treatment, and that never-takers with the treatment might be comparable to compliers with the treatment, although neither claim can be tested. If those equalities were to hold, however, then the average effect for compliers, adjusted for covariates, can be generalized to the entire population.

\subsection{Bounds}
\label{section:bounds}

To get estimates of, or do inference for, the average causal effect of the receipt of treatment in settings with non-compliance an alternative to making additional assumptions, is to focus on getting ranges of values for the estimand that are consistent with the data in a bounds or partial identification approach in a line of research associated with Manski (1990, 1996, 2003, 2013).

The simplest approach without any additional assumptions recognizes that because of the non-compliance the receipt of treatment is no longer exogenous. We can therefore analyze this as an observational study without any assumptions on the assignment process. Consider the  average difference in potential outcomes if all units are assigned to the treatment versus no-one is assigned to the treatment,
\[\tau=\frac{1}{N}\sum_{i=1}^N \Bigl( Y_i(1)-Y_i(0)\Bigr)=\oy(1)-\oy(0).\]
 To estimate this object it is useful to look at both terms separately. The first term is
\[ \oy(\ttt)=\frac{N_t}{N}\cdot \oy^\obs_\ttt+
\frac{N_c}{N}\cdot\frac{1}{\nt}\sum_{i:W_i=0} Y_i(1).\]
The last term is what is causing the problems. The data are not directly informative about this term. Let us look at the special case where the outcome is binary. In that case
\[ \oy(\ttt)\in\Bigl[\frac{N_t}{N}\cdot\oy^\obs_\ttt,
\frac{N_t}{N}\cdot \oy^\obs_\ttt+\frac{N_t}{N}
\Bigr].\]
We can do the same thing for the second term in the estimand, leading to
\[ \tau\in\Bigl[\frac{N_t}{N}\cdot\oy^\obs_\ttt-\frac{N_c}{N}\cdot \oy^\obs_\tc-\frac{N_c}{N}
,\frac{N_t}{N}\cdot \oy^\obs_\ttt+\frac{N_t}{N}-\frac{N_c}{N}\cdot \oy^\obs_\tc
\Bigr].\]
This is not a very informative range. By construction it always includes zero, so we can never be sure that the treatment has any effect on the outcome of interest.

Next, let us consider how the bounds change when we add information in the form of additional assumptions.
Under the full set of instrumental variables assumptions, that is, the monotonicity assumption and the exclusion restriction,
we can tighten the bounds substantially. To derive the bounds, and at the same time develop intuition for their value, it is useful to think of the average treatment effect as the sum of the averages over the three compliance groups, compliers, never-takers and always-takers, with shares equal to $\pi_\tc$, $\pi_\tn$, and $\pi_\ta$, respectively. Under monotonicity and the exclusion restriction the average effect for compliers is identified. For always-takers we can identify $\mme[Y_i(1)|C_i=\ta]$, but the data are uninformative about $\mme[Y_i(0)|C_i=\ta]$, so that the average effect for always-takers is bounded by
\[ \tau_\ta\in\Bigl[ \mme[Y^\obs_i|Z_i=0,W_i=1]-1,
,\mme[Y^\obs_i|Z_i=0,W_i=1]
\Bigr].\]
Similarly,
\[ \tau_\tn\in\Bigl[ -\mme[Y^\obs_i|Z_i=1,W_i=0],
,1-\mme[Y^\obs_i|Z_i=1,W_i=0]
\Bigr].\]
Combining these leads to
\[ \tau\in\Bigl[\pi_\ta\cdot \left(\mme[Y^\obs_i|Z_i=0,W_i=1]-1
\right)-\tau_\tn\cdot \mme[Y^\obs_i|Z_i=1,W_i=0]\]
\[\hskip3cm
+\left( \mme[Y^\obs_i|Z_i=1]-\mme[Y^\obs_i|Z_i=1]\right),
\]
\[\hskip2cm
\pi_\ta\cdot \mme[Y^\obs_i|Z_i=0,W_i=1]+\pi_\tn\cdot\left(1-\mme[Y^\obs_i|Z_i=1,W_i=0]\right)\]
\[\hskip3cm+\left( \mme[Y^\obs_i|Z_i=1]-\mme[Y^\obs_i|Z_i=1]\right)\Bigr].
 \]
Under these assumptions these bounds are sharp (Balke and Pearl, 1997).

\subsection{As-Treated and Per Protocol Analises}
\label{section:astreated}

There are two older methods that have sometimes been used to analyze experiments with non-compliance that rely on strong assumptions, as-treated and per-protocol analyses. See, for example, McNamee (2009) and Imbens and Rubin (2015). In an {\it as-treated} analysis units are compared by the treatment received, rather than the treatment assigned, essentially invoking an unconfoundedness assumption. Because it was the assignment that was randomized, rather than the receipt of treatment, this is not justified by the randomization. It is useful to consider in a setting where the instrumental variables assumptions, that is, the monotonicity assumption and the exclusion restriction, hold, and assess what the as-treated analysis leads to.

The estimand in an as-treated analysis is
\[ \tau^{\rm at}=\mme[Y^\obs_i|W_i=1]-\mme[Y^\obs_i|W_i=0].\]
If the monotonicity assumption holds the first term is an average of outcomes given treatment for always-takers and compliers. If the fraction of units with $Z_i=1$ is equal to $p_Z$, then we can write the first term as
\[ \mme[Y^\obs_i|W_i=1]=
\frac{\pi_\ta}{\pi_\ta+\pi_\tc\cdot P_Z}\cdot\mme[Y_i(1)|C_i=\ta]+\frac{\pi_\tc\cdot p_Z}{\pi_\ta+\pi_\tc\cdot P_Z}\cdot \mme[Y_i(1)|C_i=\tc].\]
Similarly,
\[ \mme[Y^\obs_i|W_i=0]=
\frac{\pi_\tn}{\pi_\tn+\pi_\tc\cdot (1-P_Z)}\cdot\mme[Y_i(0)|C_i=\tn]+\frac{\pi_\tc\cdot (1-p_Z)}{\pi_\tn+\pi_\tc\cdot (1-P_Z)}\cdot \mme[Y_i(0)|C_i=\tc].\]
The difference is then
\[ \mme[Y^\obs_i|W_i=1]- \mme[Y^\obs_i|W_i=0]\]
\[\hskip1cm =\mme[Y_i(1)-Y_i(0)|C_i=\tc]\]
\[\hskip2cm +\frac{\pi_\ta}{\pi_\ta+\pi_\tc\cdot P_Z}\cdot\Bigl(\mme[Y_i(1)|G_i=\ta]- \mme[Y_i(1)|C_i=\tc]\Bigr)
\]
\[\hskip2cm -\frac{\pi_\tn}{\pi_\tn+\pi_\tc\cdot (1-P_Z)}\cdot\Bigl(\mme[Y_i(0)|G_i=\tn]- \mme[Y_i(0)|C_i=\tc]\Bigr).
\]
This last two terms in expression are compared to zero in a Hausman test for the exogeneity of the treatment. The form is in general  difficult to interpret.

The second is a {per protocol} analysis, where units who do not comply with their assigned treatment are simply dropped from the analysis. Again it is instructive to see what this method is estimating under the monotonicity assumption and the exclusion restriction. In general,
\[ \tau^{\rm pp}=\mme[Y^\obs_i|W_i=1,Z_i=1]-\mme[Y^\obs_i|W_i=0,Z_i=0].\]
Similar calculations as for the as-treated analysis show that given monotoniticy and the exclusion restriction this  is equal to
\[\tau^{\rm pp} =\mme[Y_i(1)-Y_i(0)|C_i=\tc]
 +\frac{\pi_\ta}{\pi_\ta+\pi_\tc}\cdot\Bigl(\mme[Y_i(1)|G_i=\ta]- \mme[Y_i(1)|C_i=\tc]\Bigr)
\]
\[\hskip2cm -\frac{\pi_\tn}{\pi_\tn+\pi_\tc}\cdot\Bigl(\mme[Y_i(0)|G_i=\tn]- \mme[Y_i(0)|C_i=\tc]\Bigr).
\]
This expression is again dificult to interpret in general, and the analysis is not recommended.

\section{Heterogenous Treatment Effects and Pretreatment Variables}
\label{section:heterogeneity}
 
Most of the literature has focused on estimating average treatment effects for the entire sample or population. However, in many cases researchers are also interested in the presence or absence of heterogeneity in treatment effects. There are different ways to study such heterogeneity.
Here we discuss some approaches. Note that this is different from the way covariates or pretreatment variables were used in Section \ref{ran_cov}, where the focus remained on the overall average treatment effect and the presence of pretreatment variables served solely to improve precision of the estimators. In observational studies covariates also serve to make the identifying assumptions more credible.

As discussed at the outset of this chapter, a key concern with randomized experiments is external validity.
If we apply the treatment in a different setting, will the effect be the same?  Although there are many
factors that vary across settings, one common way that settings differ is that the populations of individual
units may be different.  If these differences can be captured with observable pre-treatment variables,
then it is in principle possible to address this element of external validity as in Hotz, Imbens and Mortimer (2005).  In particular, if we
obtain an estimate of the treatment effect for each potential value of the covariate vector $x$, then
we can estimate average treatment effects in any population be accounting for the differences in
distributions.  That is, given an estimate for $\tau(x)=\mme[Y_i(1)-Y_i(0)|X_i=x]$, it is straightforward to
estimate $\mme[\tau(X_i)]$ if the distribution of $X_i$ is known.

\subsection{Randomized Experiments with Pretreatment Variables}
\label{section:pretreatment}

 Traditionally researchers specified particular subpopulations based on substantive interest, and estimated average treatment effects for those subpopulations, as well as tested equality of treatment effects across these subpopulations. For example, one may be interested separately in the effect of an educational progam on girls versus boys. In such cases the analyses are straightforward. One can simply analyze the data separately by subpopulation using the methods developed in Section \ref{section:stratification_analysis}. In these cases there is often some concern that the subpopulations were selected {ex post}, so that p-values are no longer valid because of multiple testing concerns. 
For example, suppose one has a randomized experiment, with a hundred independent binary pretreatment variables that are in fact unrelated to the treatments or the outcomes. One would expect that for five of them the t-statistic for testing the null hypothesis that the average treatment effect was different by the value of that covariate was larger than 2 in absolute value, even though none of the covariates are related to the treatment effect.
Pre-analysis plans (Casey, Glennerster, and Miguel, 2012; Olken, 2015) are one approach to alleviate such concerns; another is to correct for multiple testing (List, Shaikh, and Xu, 2016).  Below we describe some recently developed alternatives that work not only when the number of covariates is small,
but also when the number is large relative to the sample size or the true underlying model of treatment effect heterogeneity may
be quite complex.

\subsection{Testing for Treatment Effect Heterogeneity}
\label{section:testingheterogeneity}

A second approach is to simply test for the presence of heterogeneity in the average treatment effect as a function of the covariates, $\tau(x)=\mme[Y_i(1)-Y_i(0)|X_i=x]$. 

One type of test considers whether there is any evidence for observable heterogeneity.  Crump, Hotz, Imbens and Mitnik (2008) develop nonparametric tests for the null hypothesis
\[ H_0:\ \tau(x)=\tau,\hskip1cm {\rm for\ all}\ x\in\mathbb{X},\]
against the alternative 
\[ H_0:\ \tau(x)\neq \tau(x'),\hskip1cm {\rm for\ some}\ x,x'\in\mathbb{X}.\]
 The Crump et al  set up uses a sequence of parametric approximations to the conditional expectation
\[ \mme[Y^\obs_i|W_i=w,X_i=x]=\beta_0'h(x)\cdot(1-w)+
\beta_1'h(x)\cdot w,\]
for vector-valued functions $h(x)$
and then tests the null hypothesis the equality $\beta_1=\beta_0$. By increasing the dimension of $h(x)$, with a suitable basis of functions, one can nonparametrically test the null hypothesis that the average treatment effect $\tau(x)$ is constant as a function of the covariates under the assumption of unconfoundedness, which is implied by randomized assignment.

A researcher might also like to understand which, if any, covariates are associated with treatment effect. A natural approach
would be to evaluate heterogeneity with respect to each covariate, one by one.  For example, each covariate could be transformed into a binary indicator for whether the value of the covariate is above or below the median, and then the researcher could test the
hypothesis that the treatment effect is higher when the covariate is high than when it is low.  Conducting a large
number of hypothesis tests raises issues of multiple testing, and confidence intervals should be be corrected to account for this.  However, standard approaches (e.g. the Bonferroni correction) assume that each test is independent, and thus may be 
overly conservative in an environment where many covariates are correlated with one another (which will imply that the test
statistics are also correlated with one another).  List, Shaikh, and Xu (2016) propose a computationally feasible approach to 
the multiple testing problem in this context.  The approach uses bootstrapping, and it accounts for correlation among test
statistics. One challenge with this approach is that the researcher must pre-specify the 
set of hypothesis tests to conduct; thus, it is hard to explore all possible interactions among covariates and all possible ways to
discretize them.  In the next section, we consider methods that explore more complex forms of heterogeneity. 

\subsection{Estimating Treatment Effect Heterogeneity}
\label{section:exploringheterogeneity}

There are several possible approaches for exploring treatment effect heterogeneity.  The first
is to specify a parametric model of treatment effect heterogeneity (as in (\ref{interactions})) and report the estimates.  For example,
one simple approach would be to specify a regression of the outcome on an indicator for treatment 
status as well as interactions of the indicator with the treatment indicator.  With a small number of
covariates relative to the sample size, all linear interactions with the treatment indicator could be
considered, partially alleviating concerns about multiple testing.  Below we discuss generalizations
of this idea to regularized regression (e.g. LASSO) where a systematic method is used to select covariates.

A second approach is to construct a fully nonparametric estimator for $\tau(x)$.  We will develop this
approach further below; with sufficiently large datasets and a relatively small number of covariates,
this approach can be effective, and recent work building on techniques from machine learning (Wager
and Athey, 2015) has 
lead to improvements in how many covariates can be handled without sacrificing coverage of confidence intervals.  For the case where
there may be many covariates relative to the sample size, a third approach proposed by Athey and
Imbens (2015) uses the data to select a set of subgroups (a ``partition'' of the covariate space) such
that treatment effect heterogeneity across subgroups is maximized in a particular sense.  

Whether
a fully non-parametric approach or an approach based on subgroups is preferred may be partially
determined by the constraints of the data; valid confidence intervals may not be available (at least
with existing methods) with too many covariates relative to sample size.  But even if both methods
are potentially feasible, it may be desirable to learn about subgroups rather than a fully nonparametric
estimate of $\tau(x)$ if the results of the experiment will be used in a context where people with limited processing capability/memory will make decisions based on the experiment.  For example, doctors might
use a simple flowchart to determine which patients should be prescribed a drug.  Results about subgroups may also be
more easily interpretable by researchers.  

Relative to testing all covariates one by one, an approach that selects a single
partition of the covariate space will not in general discover all heterogeneity that exists, since the algorithm will focus on the covariates with the biggest impact
to the exclusion of others.  In addition, in the process of constructing a partition, once we have divided the data into two groups
according to the value of one covariate, further divisions will be considered
on subsamples of the data, reducing the power available to test heterogeneity in additional covariates.  
Thus, constructing a single partition does not answer the question of which covariates are associated with heterogeneity;
rather, it identifies a particular way to divide the data into meaningful groups.  If a researcher wanted to explore all covariates, while maintaining a data-driven
approach to how to discretize them, an approach would be to
construct distinct partitions that restrict attention to one covariate at the time.  For interactions, one could consider small subsets of covariates. If the
results of such an exercise were reported in terms of which covariates are associated with significant heterogeneity, multiple testing corrections would be warranted.  The approach of List, Shaikh, and Xu (2016) works for an arbitrary set of null hypotheses,
so the researcher could generate a long list of hypotheses using the causal tree approach restricted to different subsets of 
covariates, and then test them with a correction for multiple testing. Since in datasets with many covariates, there are often
many ways to describe what are essentially the same subgroups, we expect a lot of correlation in test statistics, reducing the
magnitude of the correction for multiple hypothesis testing.

We begin by describing the third approach, where we construct a partition of the covariate space, and then return to the second and first approaches.

\subsubsection{Data-driven Subgroup Analysis: Recursive Partitioning for Treatment Effects}
\label{section:trees}

Athey and Imbens (2016) develop a method for exploring heterogeneity in treatment effects without having to prespecify the form of the heterogeneity, and without having to worry about multiple testing. Their approach builds on ``regression tree'' or ``recursive partitioning'' methods, where the sample is partitioned in a number of subgroups, defined by the region of the covariate space each unit belongs to. 
The data is used to determine which partition produces subgroups that differ
the most in terms of treatment effects. The method avoids introducing biases in the estimated average treatment effects and allows for valid confidence intervals using ``sample splitting,'' or ``honest'' estimation. The idea of sample splitting to control significance levels goes back a long way in statistics; see, e.g. Cox (1975),
or for a more recent discussion, see Fithian, Sun, and Taylor (2015).  In a sample splitting approach, in a first step, one sample is used to select the partition, while in a second step
an independent sample is used to estimate treatment effects and construct confidence intervals for each subgroup (separately) given the partition from the first step.  The output of the method is a set of subgroups, selected
to optimize for treatment effect heterogeneity (to minimize expected mean-squared error of treatment effects), together
with treatment effect estimates and standard errors for each subgroup.

Let us illustrate some of the issues in a simple case to develop  more intuition. Suppose we consider only a single split of the covariate space, in a setting with a substantial number of covariates. We specify a criterion that determines whether one split (that is, a combination of a choice of the covariate and a threshold) is better than another. We return to the choice of criterion below. Given a criterion, we select the covariate and threshold that maximize the citerion. If we estimate the average treatment effect on the two subsamples using the same sample, the fact that this particular split led to a high value of the criterion would often imply that the average treatment effect estimate is biased. Athey and Imbens (2016) therefore suggest, in what they call an {honest} approach, to estimate the treatment effects on a separate sample. The implication is that the treatment effect estimates are unbiased on the two subsamples, and the corresponding confidence intervals are valid, even in settings with a large number of pretreatment variables or covariates.

A key issue is the choice of criterion. In principle one would like to split in order to obtain more precise estimates of the average treatment effects. A complicating factor is that the standard criterion for splitting optimized for prediction rely on observing the outcome whose expectation one wants to estimate. That is not the case here because the unit-level treatment effect is not observed. There have been various suggestions in the literature to deal with this. 
One simple solution is to transform the outcome from $Y^\obs_i$ to
\[ Y^*_i=Y^\obs_i\cdot \frac{W_i-p}{p\cdot (1-p)}.\]
This transformed outcome has the property that $\mme[Y^*_i|X_i=x]=\tau(x)=\mme[Y_i(1)-Y_i(0)|X_i=x]$ so that standard methods for recursive partitioning based on prediction apply (see Weisberg and Pontes (2015), Athey and Imbens (2016)).
Su et al (2009) suggest using test statistics for the null hypothesis that the average treatment effect in the two subsamples are equal to zero. Zeileis, Hothorn, and Hornick (2008) suggest using model fit, where the model corresponds to a linear regression model in the partitions with an intercept and a binary treatment indicator.\footnote{Neither of those two papers consider honest estimation, nor do they establish bias and consistency properties
of the estimators.} Athey and Imbens (2016) show that neither of the two criteria is optimal, and derive a new criterion that focuses directly on the expected squared error of the treatment effect estimator, and which turns out to depend both on the t-statistic and on the fit measures.  The criterion is further modified to anticipate honest estimation, that is, to anticipate that the treatment effects will
be re-estimated on an independent sample after the subgroups are selected.  This modification ends up penalizing the expected
variance of subgroup estimates; for example, if subgroups are too small, the variance of treatment effect estimates will be large.  It also
rewards splits for covariates that explain outcomes but not treatment effect heterogeneity, to the extent that controlling for such
covariates enables a lower-variance estimate of the treatment effect.

Other related approaches 
include Wager and Walther (2015), who
discuss corrections to confidence intervals (widening the confidence intervals by a factor) as an alternative to sample splitting; however, since confidence intervals need to be inflated fairly substantially, it is not clear whether there is a wide range of conditions where it improves on sample splitting.  Relative to the approach of List, Shaikh, and Xu (2016) discussed above, the methods in this section focus on deriving a single partition, rather than considering heterogeneity one covariate at the time with pre-specified discretizations of the covariates; the approaches in this section will have the advantage of exploring interaction effects and using the data to determine
a meaningful partition in terms of mean-squared error of treatment effects.

\subsubsection{Non-Parametric Estimation of Treatment Effect Heterogeneity}

There are (at least) four possible goals for using non-parametric estimation to estimate heterogeneous treatment effects.  The
first is descriptive: the researcher can gain insight about what types of units have the highest and lowest treatment effects,
as well as visualize comparative statics results, all without imposing a prior restrictions.  The second, disucssed earlier, is that
the researcher wishes to estimate the impact of applying the treatment in a setting with a different distribution of units.  A third
is that the researcher wishes to derive a personalized policy recommendation.  A fourth
is that the researcher wishes to test hypotheses and construct confidence intervals.  If confidence intervals are desired, the set
of potential methods is quite small.  For optimal policy evaluation, a Bayesian framework may have some advantages, since
it is natural to incorporate the uncertainty and risk involved in alternative policy assignments.  For description or estimation where
confidence intervals are not important, there are a wide variety of approaches.

Classical non-parametric approaches to treatment effect heterogeneity would include K-nearest neighbor  matching  and
kernel estimation (H\"ardle, 2002).  In the case of $K-$nearest neighbor matching, for any $x$ we can construct an estimate of the treatment effect at that
$x$ by averaging the outcomes of the $K$ nearest neighbors that were treated, and subtracting the average outcomes of
the $K$ nearest neighbors that were control observations.  Kernel estimation does something similar, but uses a smooth weighting
function rather than uniformly weighting nearby neighbors and giving 0 weight to neighbors that are farther away.  In both
cases, distance is measured using Euclidean distance for the covariate vector.  These methods can work well and provide satisfactory
coverage of confidence intervals with one or two covariates, but performance deteriorates quickly after that.  The output of the
nonparametric estimator is a treatment effect for an arbitrary $x$.  The estimates generally must be further summarized or visualized
since the model produces a distinct prediction for each $x$.

A key problem with kernels and nearest neighbor matching is
that all covariates are treated symmetrically; if one unit is close to another in 20 dimensions, the units are probably not particularly
similar in any given dimension.  We would ideally like to prioritize dimensions that are most important for heterogeneous treatment
effects, as is done in many machine learning methods, including the highly successful random forest algorithm.  Unfortunately, many popular machine learning methods that use the data to select covariates may be bias-dominated asymptotically (including the standard
random forest).
Recently, Wager and Athey (2015) propose a modified version of the random forest algorithm that produces treatment effect estimates that can be shown to be asymptotically
normal and centered on the true value of the treatment effect, and they propose a consistent estimator for the asymptotic variance.  The method averages over many ``trees'' of the
form developed in Athey and Imbens (2016); the trees differ from one another because different subsamples are used for each tree, and in addition there is some randomization
in the choice of which covariates to split on.  Each tree is ``honest,'' in that one subsample is used to determine a partition and an
independent subsample is used to estimate treatment effects within the leaves.  Unlike the case of a single tree, no data is ``wasted''
because each observation is used to determine the partition in some trees and used to estimate treatment effects in other trees,
and subsampling is already an inherent part of the method.  The method can be understood as a generalization of kernels and
nearest neighbor matching methods, in that the estimated treatment effect at $x$ is the difference between a weighted average of nearby 
treated units and nearby control units; but the choice of what dimenions are important for measuring distance is determined by
the data.  In simulations, this method can obtain nominal coverage with more covariates than K-Nearest Neighbour matching or kernel methods, while simultaneously
producing much more accurate estimates of treatment effects.  However, this method also eventually becomes bias-dominated
when the number of covariates grows.  It is much more robust to irrelevant covariates than kernels or nearest neighbor matching.

Another approach to the problem is to divide the training data by treatment status, and apply supervised
learning methods to each group separately.  For example,
Foster, Taylor, and Ruberg (2011) use random forests to estimate the effect of covariates on outcomes in treated and control groups.  They then take the difference in predictions as data and project treatment effects onto units' attributes using regression or classification trees.  The approach of Wager and Athey (2015) can potentially gain efficiency by directly estimating heterogeneity in causal effects, and further the off-the-shelf random forest estimator does not have established statistical properties (so confidence intervals are
not available).

Taking the Bayesian perspective, Green and Kern (2012) and Hill (2011) have proposed the use of forest-based algorithms for estimating heterogeneous treatment effects. These papers use the Bayesian additive regression tree  method of Chipman et al (2010), and report posterior credible intervals obtained by Markov-chain Monte Carlo  sampling based on a convenience prior. Although Bayesian regression trees are often successful in practice, there are currently no results guaranteeing posterior concentration around the true conditional mean function, or convergence of the Markov-Chain-Monte-Carlo sampler in polynomial time.  In a related paper, Taddy, Gardner, Chen and Draper (2014) use Bayesian nonparametric methods with Dirichlet priors to flexibly estimate the data-generating process, and then project the estimates of heterogeneous treatment effects down onto the feature space using regularization methods or regression trees to get low-dimensional summaries of the heterogeneity; but again, asymptotic properties are unknown.

\subsubsection{Treatment Effect Heterogeneity Using Regularized Regression}

Imai and Ratkovic (2013), Signorovitch (2007), Tian et al (2014),  and Weisberg and Pontes (2015) develop lasso-like methods for causal inference and treatment effect heterogeneity in a setting where there are potentially a large number of covariates, so that regularization methods to discover which covariates
are important.  When the treatment effect interactions of interest have low dimension (that is, a small number
of covariates have important interactions with the treatment), valid confidence intervals can be derived (without using sample
splitting as described above); see, e.g., Chernozhukov, Hansen, and Spindler (2015) and references therein.  These methods require that the true underlying model is (at least approximately) ``sparse'': the number of observations must be large relative to the number of covariates
(and their interactions) that have an important effect on the outcome and on treatment effect heterogeneity.  Some of the methods
(e.g. Tian et al (2014)) propose modeling heterogeneity in the treatment and control groups separately, and then taking the difference;
this can be inefficient if the covariates that affect the level of outcomes are distinct from those that affect treatment effect
heterogeneity.  An alternative approach is to incorporate interactions of the treatment with covariates as covariates, and then allow LASSO 
to select which covariates are important.  Interaction terms can be prioritized over terms that do not include treatment effect
interactions through weighting. 

\subsubsection{Comparison of Methods}

Although the LASSO based methods require more a priori restrictions on sparsity than the random forest methods, both types of
methods will lose nominal coverage rates if the models become too complex.  The LASSO methods have some advantages with
datasets where there are linear or polynomial relationships between covariates and outcomes; random forest methods do not
parsimoniously estimate linear relationships and use them for extrapolation, but are more localized.  The random forest methods
are well-designed to capture complex, multi-dimensional interactions among covariates, or highly nonlinear interactions.  LASSO
has the advantage that the final output is a regression, which may be more familiar to researchers in some disciplines; however,
it is important to remember that the conditions the justify the standard errors are much more stringent when the model
selection was carried out on the same data that is used for estimation.  If valid
confidence intervals are the first priority in an environment where the model is not known to be sparse and there are many covariates,
the recursive partitioning approach provides confidence intervals that do not deteriorate (at all) as the number of covariates
grow.  What suffers, instead, is the mean-squared error of the predictions of treatment effects.

Another point of comparison between the regression-based methods and tree-based methods (including random forests) 
relates to our earlier discussions of randomization-based inference versus sampling-based inference.  Tree-based methods
construct estimates by dividing the sample into subgroups and calculating sample averages within the groups; thus,
the estimates and associated inference can be justified by random assignment of the treatment. In contrast, regression-based
approaches require additional assumptions.

\subsubsection{Relationship to Optimal Policy Estimation}

The problem of estimating heterogeneous treatment effects is closely related to the problem of 
estimating, as a function of the covariates, what the optimal policy is.  Heuristically, with a binary
treatment, we would want to assign an individual with covariates $x$ to a treatment if $\tau(x)>0$.  However,
the optimal policy literature addresses additional issues that might arise when there are multiple potential treatments,
as well as when the loss function may be nonlinear (so that there is, for example, a mean-variance tradeoff between 
different policies).  More broadly, the criterion used in estimation may be modified to account for the goal of policy
estimation; when regularization approaches are used to penalize model complexity, the methods may de-prioritize
discovering heterogeneity that is not relevant for selecting an optimal policy.  For example, if a treatment clearly dominates
another for some parts of the covariate space, understanding heterogeneity in the magnitude of the treatment's advantage
may not be important in those regions.

Much of the policy estimation literature takes a Bayesian perspective; this allows the researcher to evaluate welfare and
to incorporate risk aversion in the loss function in an environment where there is uncertainty about the effects of the policy.

In the machine learning literature, Beygelzimer and Langford (2009) and Dudik, Langford and Li (2011) discuss procedures for transforming outcomes that enable off-the-shelf loss minimization methods to be used for optimal treatment policy estimation.
In the econometrics literature, Graham, Imbens and Ridder (2014), Dehejia (2005), Hirano and Porter (2009), Manski (2004), and Bhattacharya and Dupas (2012) estimate parametric or semi-parametric models for optimal policies, relying on regularization for covariate selection in the case of Bhattacharya and Dupas (2012). 
See also  Banerjee, Chassang and Snowberg (2015).

\section{Experiments in Settings with Interactions}
\label{section:interactions}

In this section we discuss the analysis of randomized experiments in settings with interference between units. Such interference may take different forms. There may be spillovers from the treatment assigned to one unit to other units. A classic example of that is that of agricultural experiments where fertilize applied to one plot of land may leach over to other plots and thus affect outcomes in plots assigned to different treatments. It may also take the form of active and deliberate interactions between individuals, for example in educational settings, where exposing one student to a  new program may well affect the outcomes for students the first student is friends with. 

There are many different versions of these problems, and many different estimands. 
An important theoretical paper in an observational setting is Manski (1993) who introduced terminology to distinguish between contextual effects, exogenous effects, and endogenous effects. Contextual effects arise when individuals are exposed to similar environmental stimula as their peers. Exogenous effects refer to effects from fixed characteristics of an individual's peers. Endogenous effects in Manski's terminology refer to direct causal effects of the behavior of the peers of an individual.

The interactions may be a nuisance that affects the ability to do inference, with the interest in the overall average effect, or the interactions may be of primary interest to the researcher. This is an active area of research, with many different approaches, where it it not clear what will ultimately be the most useful results for empirical work.
In fact, some of the most interesting work has been empirical.

\subsection{Empirical Work on Interactions}

Here we discuss some of the questions raised in empirical work on interactions. These provide some of the background for the discussions of the theoretical work by suggesting particular questions and settings where these questions are of interest. There are a number of different settings. In some cases the peer group composition is randomized, and in other cases treatments are randomized. An example of the first case is Sacerdote (2001) where individual students are randomly assigned to dorm rooms and thus matched to a roommate. An example of the second is Duflo and Saez (2003) where individuals in possibly endogenously formed groups are randomly assigned to treatments, with the treatments clustered at the group level.

Miguel and Kremer (2004) were interested in the effects of deworming programs on children's educational outcomes. There are obviously direct effects of deworming on the outcomes for individuals who are exposed to these programs, but just as in the case of infectious diseases in general, there may be externalities for individuals not exposed to the program if individuals they interact with are exposed. Miguel and Kremer find evidence of substantial externalities.

Crepon et al (2013) were interested in the effects of labor market training programs. They were concerned about interactions between individuals through the labor market. Part of the effect of providing training to an unemployed individual may be that this individual becomes more attractive to an employer relative to an untrained individual. If, however, the total number of vacancies is not affected by the presence of more trained individuals, the overall effect may be zero even if the trained individuals are more likely to be employed than the individuals in the control group. Cr\'epon et al studied this by randomizing individuals to training programs in a number of labor markets. They varied the marginal rate at which the individuals were assigned to the training program between the labor markets. They then compared the difference  in average outcomes by treatment status within the labor markets, across the different labor markets. In the absence of interactions, here in the form of equilibrium effects, the average treatment effects should not vary by the marginal treatment rate. Evidence that the average treatment effects were higher when the marginal treatment rate was lower suggests that part of the treatment effect was based on redistributing jobs from control individuals to trained individuals.

Sacerdote (2001) studies the effect of roommates on an individual's behavior in college. He exploits the random assignment of incoming students to dorm rooms at Dartmouth (after taking account of some characteristics such as smoking behavior). The treatment can be thought of here as having a roommate of a particular type, such as a roommate with a relatively high or low level of high school  achievement. If roommates are randomly assigned, then finding that individuals with high achieving roommates have outcomes that are systematically different from those of individuals with low achieving roommates is evidence of causal  interactions between the roommates.

Carrell, Sacerdote and West (2013) analyze data from the US Air Force Academy. They control the assignment of incoming students to squadrons to manipulate the distribution of characteristics of fellow squadron  students that an incoming student is faced with. They find that the outcomes for students vary systematically with this distribution of fellow student characteristics, which is evidence of causal effects of interactions.

\subsection{The Analysis of Randomized Experiments with Interactions in Subpopulations}
\label{section:groups}

One important special case of interference assumes the population can be partitioned into groups or clusters, with the interactions limited to units within the same cluster. This is a case studied by, among others, Manski (2013) and
Hudgens and Halloran (2008), and Liu and Hudgens (2013).
Ugander,  Karrer, Backstrom and Kleinberg (2013) discuss graph cutting methods in general network settings to generate partitions of the basic network where such an assumption holds, at least approximately.
Hudgens and Halloran define in this setting direct, indirect, total and overall causal effects, and consider a two-stage randomized design where in the first stage clusters are selected randomly and in the second stage units within the clusters are randomly assigned. The direct causal effect for a particular unit corresponds to the difference between potential outcomes where only the treatment effect for that unit is changed, and all other treatment are kept fixed. Indirect effects correspond to causal effects of changes in the assignments for other units in the same group, keeping fixed the own assignment. The total effect combines the direct and indirect effects. Finally, the 
overall effect in the Hudgens and Halloran framework is the average effect for a cluster or group, compared to the baseline where the entire group receives the control treatment.

Hudgens and Halloran also stress the widely used assumption that for unit $i$ it matters only what fraction of the other units in their group are treated, not the identity of the treated units. Without such an assumption the proliferation of indirect treatment effects makes it difficult to obtain unbiased estimators for any of them. This assumption is often made, sometimes implicitly, in empirical work in this area.

They consider designs where the marginal rate of treatment varies across groups. In the first stage of the assignment the groups are randomly assigned to different treatment rates, followed by a stage in which the units are randomly assigned to the treatment.

\subsection{The Analysis of Randomized Experiments with Interactions in Networks}
\label{section:networks}

Here we look at a general network setting where the population of units is not necessarily partitioned into mutually exclusive groups. With $N$ individuals in the population of interest we have a network characterized by an $N\times N$ adjacency matrix $G$, with $G_{ij}\in\{0,1\}$ a binary indicator for the event that units $i$ and $j$ are connected. The matrix $G$ is symmetric with all diagonal elements equal to zero. The question here is what we can learn about presence interaction effects by conducting a randomized experiment on this single network, with a binary treatment. Unlike the Manski (1993) and Hudgens and Halloran (2008)  setting, we have only a single network, but the network is richer in the sense that it need not be the case that friends of friends are also friends themselves. The type of questions Athey, Eckles and Imbens (2015) are interested in are, for example, whether there is evidence that changing the treatment for friends affects an individual's outcome, or whether manipulating treatments for friends of an individual's friends changes there outcome. They do so by focusing on exact tests to avoid the reliance on large sample approximations, which can be difficult to derive in settings with a single network.  (There is not even a clear answer to the question of what it means for a network to grow in size; specifying this would require the researcher to specify what it means for the network to grow, in terms of new links 
for new units.) 

Let us focus in this discussion on the two main hypotheses Athey, Eckles and Imbens (2015) consider. First, the null hypothesis of no interactions whatsoever, that is the null hypothesis that changing the treatment status for friends does not change an individual's outcome, also considered in Aronow (2012), and second, the null hypothesis that the treatment of a friend of a friend does not have a causal effect on an individual's outcome. Like Liu and  Hudgens (2013) and Athey, Eckles and Imbens (2015), they consider randomization inference.

We focus on the setting where in the population treatments are completely randomly assigned. The network itself is analyzed as given. Initially let us focus on the null hypothesis of no interactions whatsoever.
Athey, Eckles and Imbens (2015) introduce the notion of an articifial experiment. The idea is to select a number of units from the original population, whom they call the {focal} units. Given these focal units they define a test statistic in terms of the outcomes for these focal units, say the correlation between outcomes and the fraction of treated friends. They look at the distribution of this statistic, induced by randomizing the treatments only for the non-focal or {auxiliary} units. Under the null hypothesis of no treatment effects whatsoever, changing the treatment status for auxiliary units would not change the value of the outcomes for the focal units.

For the second null hypothesis, that friends of friends have no effect, they again consider a subset of the units to be focal.
A second subset of units are termed ``buffer'' units: these are the friends of the focal unit.  If we allow that friends can have
an impact on focal units, then their treatments cannot be randomized in the artificial experiment designed to test the
impact of friends of friends.  The complement of focal and buffer units, termed the auxiliary units, are the units whose treatments
are randomized in the artificial experiment.  The randomization distribution over the treatment assignments of auxiliary units induces
a distribution on the test statistic, and this approach thus enables the researcher to test the hypothesis that friends of friends
have no effect, without placing any restrictions on direct effects or the effect of friends.

Athey, Eckles and Imbens (2015) also consider richer hypotheses, such as hypotheses about what types of link definitions correspond
to meaningful peer effects; they propose a test of the hypothesis that a sparser definition of the network is sufficient to capture
relationships for which treating a friend influences a unit.

Aronow and Samii (2013) study estimation in this general network setting. They assume that there is a structure on the treatment effects so that only a limited number of unit-level treatment assignments have a non-zero causal effect on the outcome for unit $i$. The group structure that Hudgens and Halloran (2008) use is a special case where it is only the treatments for units in the same group as unit $i$ can have non-zero effects on the outcome for unit $i$.


\section{Conclusion}
\label{section:conclusion}

In this chapter we discuss statistical methods for analyzing data from randomized experiments. We focus primarily on randomization-based, rather than model-based methods, starting with classic methods developed by Fisher and Neyman, up to recent work on non-compliance, clustering and methods for identifying treatment effect heterogeneity, as well as experiments in settings with interference.

\newpage

\centerline{\sc References}

\begin{description}

\item[]{\small\textsc{Abadie, A., J. Angrist, and G. Imbens}, (2002),
``Instrumental Variables Estimates of the Effect of Subsidized Training on the Quantiles of Trainee Earnings'', {\it Econometrica}, Vol. 70(1): 91-117.}

\item[]{\small\textsc{Abadie, A., S. Athey, G. Imbens, and J. Wooldridge}, (2014),
``Finite Population Causal Standard Errors'', NBER Working Paper 20325.}

\item[]{\small\textsc{Abadie, A., S. Athey, G. Imbens, and J. Wooldridge}, (2016),
``Clustering as a Design Problem'', Unpublished Working Paper.}

\item[] 
{\small\textsc{Allcott, H.,} (2015), ``Site Selection Bias in Program Evaluation,'' {\it Quarterly  Journal of Economics}, 1117-1165.}

\item[] 
{\small\textsc{Altman, D.,} (1991), {\it Practical Statistics for Medical Research}, Chapman and Hall/CRC. }

\item {\small \textsc{Angrist, J., } (2004),
``Treatment Effect Heterogeneity in Theory and Practice,'' , {\it The Economic Journal} 114(494),
C52-C83. }

\item[] 
{\small\textsc{J. Angrist, G. Kuersteiner}, (2011), 
\textquotedblleft
Causal Effects of Monetary Shocks: Semiparametric Conditional Independence Tests with a Multinomial Propensity Score,\textquotedblright {\it
Review of Economics and Statistics},  Vol. 93, No. 3:725-747. }

\item {\small \textsc{Angrist, J., and S. Pischke,} (2009), \textit{Mostly
Harmless Econometrics}, Princeton University Press, Princeton, NJ.}

\item[] 
{\small\textsc{Anscombe, F.J.} (1948).  The validity of comparative experiments.  Journal of the Royal Statistical Society, Series A, 61, 181-211.}

\item[]{\small \textsc{Aronow, P.,} (2012),
``A general method for detecting interference between units in randomized experiments,''
{\it 
Sociological Methods \& Research}, Vol. 41(1): 3-16.}

\item[]{\small \textsc{Aronow, P., and C. Samii}, (2013),
``Estimating Average Causal Effects Under Interference Between Units'', arXiv:1305.6156(v1).}

\item[]{\small\textsc{Athey, S., D. Eckles, and G. Imbens}, (2015),
``Exact P-values for Network Interference'', NBER Working Paper 21313.}

\item[]{\small\textsc{Athey, S., and G. Imbens}, (2016),
``Recursive Partitioning for Heterogeneous Causal Effects,''
 arXiv:1504.01132, forthcoming, {\it Proceedings of the National Academy of Sciences.}} 

\item[]{\small\textsc{Baker, S.}, (2000),
``Analyzing a Randomized Cancer Prevention Trial with a Missing Binary Outcome, an Auxiliary Variable, and All-or-None Compliance,''
 {\it The Journal of the American Statistical Association}, 95(449): 43-50. 
}

\item[]{\small\textsc{Baker, S.G. Kramer, B.S., and Lindeman, K.S.}, (2016), ``Latent class instrumental variables: a clinical and biostatistical perspective'' {\it Statistics in Medicine}, Vol 35(1): 147-160.}

\item[]{\small\textsc{Baker SG, and Lindeman KS}, (1994), ``The paired availability design: a proposal for evaluating epidural analgesia during labor,'' {\it  Statistics
in Medicine}, Vol. 13:2269–2278.}

\item[]{\small\textsc{Balke, A., and J. Pearl}, (1997),
``Bounds on Treatment Effects from Studies with Imperfect Compliance,''
 {\it The Journal of the American Statistical Association}, 92(439): 1171-1176. 
}

\item[] {\small \textsc{Banerjee, A., S. Chassang, and E. Snowberg}, (2016),
``decision Theoretic Approaches to Experimental Design and External Validity'', 
{\it Handbook of Development Economics}, Banerjee and Duflo (eds.), Elseviers, North Holland.}

\item[] {\small \textsc{Banerjee, A., and E. Duflo,} (2009), ``The Experimental Approach to Development Economics'', {\it Annual Review of Economics}, Vol. 1: 151-178.}

\item[] {\small \textsc{Bareinboim, E., S. Lee, V. Honavar, and J. Pearl,} (2013), ``Causal Transportability from Multiple Environments
		  with Limited Experiments'', {\it Advances in Neural Information Processing Systems 26 
		  ({NIPS} Proceedings)}, 136-144 .}


\item[]{\small\textsc{Barnard, J., J. Du, J. Hill, and D. Rubin}, (1998),
``A Broader Template for Analyzing Broken Randomized Experiments,''
 {\it Sociological Methods \& Research}, Vol. 27, 285-317. 
}

\item[]{\small\textsc{Bertanha, M., and G. Imbens}, (2014),
``External Validity in Fuzzy 
 Regression Discontinuity Designs,'' NBER working paper 20773.}

\item[] {\small \textsc{Bertrand, M., and E. Duflo}, (2016), ``Field Experiments on Discrimination'', NBER Working Paper 22014.}

\item[]{\small\textsc{Beygelzimer, A.  and J. Langford}, (2009),  ``The Offset Tree for Learning with Partial Labels'', \\  http://arxiv.org/pdf/0812.4044v2.pdf.}

\item[]{\small\textsc{Bhattacharya, Debopam and Dupas, Pascaline}, ``Inferring welfare maximizing treatment assignment under budget constraints',' {\it Journal of Econometrics}, 167(1), 168-196.}

\item[]{\small\textsc{Bitler, M., J. Gelbach,  and H. Hoynes} (2002)
``What Mean Impacts Miss: Distributional Effects of Welfare Reform 
Experiments,''  
{\it American Economic Review},  Vol. 96(4): 988-1012. 
}

\item[]{\small\textsc{Bloom, H.}, (1984),
``Accounting for No-Shows in Experimental Evaluation Designs,''
{\it Evaluation Review}, 8: 225-246. 
}

\item {\small \textsc{Bruhn, M., and D. McKenzie}, (2009), ``In Pursuit of Balance: Randomization in Practice in Development Field Experiments,'' {\it American Economic Journal: Applied Economics}, Vol. 1(4): 200-232.}

\item[]{\small \textsc{Carrell, S.,
B. Sacerdote, and
J. West} (2013),
``From Natural Variation to Optimal Policy? The Importance of Endogenous Peer Group Formation'' {\it Econometrica}, 81(3): 855-882.}

\item[]{\small\textsc{
Casey, K.,
R. Glennerster, and
E. Miguel}, (2012),
``Reshaping Institutions: Evidence on Aid Impacts Using a Preanalysis Plan,''
{\it Quarterly Journal of Economics}, 755�1812.
}

\item[]{\small\textsc{Chernozhukov, V., and C. Hansen}, (2005),
``An IV Model of Quantile Treatment Effects'', {\it Econometrica}, Vol. 73(1): 245-261.}

\item[]{\small\textsc{Chernozhukov, V., C. Hansen, and M. Spindler}, (2015),
``Post-Selection and Post-Regularization Inference in Linear Models with Many Controls and Instruments'', {\it American Economics Review}, Papers and Proceedings.}

\item[]{\small\textsc{Chipman, H., George, E., and R. McCulloch,} (2010), {\it BART: Bayesian additive regression trees}, The Annals of Applied Statistics, 266--298, 4(1).}

\item {\small \textsc{Cochran, W.}, (1972), ``Observational Studies,'' in {\it Statistical
Papers in Honor of George W. Snedecor,} ed. T.A. Bancroft, 1972, Iowa State University
Press, pp. 77-90, reprinted in {\it Observational Studies}, 2015.}

\item {\small \textsc{Cochran, W., and G. Cox}, (1957), {\it Experimental Design}, Wiley Classics Library.}

\item {\small \textsc{Cohen, J.}, (1988), {\it Statistical Power for the Behavioral Sciences}, second edition,  .}

\item[] 
{\small\textsc{Cook, T., and D. DeMets} (2008), {\it Introduction to Statistical Methods for Clinical Trials}, Chapman and Hall/CRC.}

\item[] 
{\small\textsc{Cox, D.} (1956), ``A Note on Weighted Randomization,'' {\it The Annals of Mathematical Statistics}, Vol. 27(4): 1144-1151.} 

\item[]
{\small\textsc{Cox, D.} (1975), ``A note on data-splitting for the evaluation of significance levels.'' {\it Biometrika}, 62(2): 441–444, 1975.}

\item[] 
{\small\textsc{Cox, D.,} (1992),  ``Causality: Some Statistical Aspects,'' {\it   Journal of the Royal Statistical Society}, Series A, Vol. 155, 291-301.
}

\item[]{\small \textsc{Cox, D., and N. Reid}, (2000),
{\it The Theory of the Design of Experiments}, Chapman and Hall/CRC, Boca Raton, Florida. }

\item[]{\small \textsc{Crepon, B., E. Duflo, M. Gurgand, R. Rathelot, and P. Zamora}, (2013),
``Do Labor Market Policies have Displacement Effects? Evidence
from a Clustered Randomized Experiment,'' {\it The Quarterly
Journal of Economics} 128, no. 2 (April 24, 2013): 531-580.}

\item {\small \textsc{Crump, R., V. J. Hotz, V. J., G. Imbens, and O.Mitnik}%
, (2008), \textquotedblleft Nonparametric Tests for Treatment Effect
Heterogeneity,\textquotedblright  {\it Review of Economics and
Statistics}, 90(3): 389-405.}

\item[]{\small\textsc{Cuzick, J., R. Edwards,  and N. Segnan}, (1997),
``Adjusting for non-compliance
and contamination in randomized clinical trials,''
{\it Statistics in Medicine,} Vol. 16: 1017-1039. }

\item[] 
{\small\textsc{Davies, O.,} (1954), {\it The Design and Analysis of Industrial Experiments}, Edinburgh, Oliver and Boyd.}

\item[]{\small\textsc{Deaton, A.}, (2010),
``Instruments, Randomization, and Learning about Development,'' {\it Journal of Economic Literature}, 424-455. }

\item[]{\small\textsc{Dehejia, R.}, (2005), ``Program Evaluation as a Decision Problem,'' {\it Journal of Econometrics}, 125(1), 141-173.}

\item[]{\small \textsc{Diehr, P., D. Martin, T. Koepsell, and A. Cheadle}, (1995),
\textquotedblleft Breaking the Matches in a Paired $t$-Test for Community Interventions when the Number of Pairs is Small,\textquotedblright\ {\it Statistics in Medicine} Vol 14 1491-1504.
}

\item[]{\small \textsc{Donner, A.}, (1987),
\textquotedblleft Statistical Methodology for Paired Cluster Designs,\textquotedblright\ {\it American Journal of Epidemiology,} Vol 126(5), 972-979. }

\item[]{\small \textsc{Doksum, K.}, (1974), ``Empirical Probability Plots and Statistical Inference for Nonlinear Models in the Two-Sample Case,''  {\it Annals of Statistics}, 2, 267-277.}

\item[]{\small \textsc{Dudik, M.,  J. Langford and L. Li}, (2011), ``Doubly Robust Policy Evaluation and Learning'', {\it Proceedings of the 28th International Conference on Machine Learning} (ICML-11).}

\item[]{\small \textsc{Duflo, E., R. Glennerster, and M. Kremer}, (2006), ``Using Randomization in Development Economics Research: A
Toolkit,''  {\it Handbook of Development Economics}, Elseviers.}

\item {\small \textsc{Duflo, Esther, Rema Hanna, and Stehpen Ryan}, (2012), 
\textquotedblleft  Incentives Work: Getting Teachers to Come to School,\textquotedblright, {\it American Economic Review}, 102(4): 1241-78.}

\item[]{\small\textsc{Duflo, E., and E. Saez}, (2003),
``The Role of Information and Social Interactions in Retirement Decisions: Evidence from a Randomized Experiment,'' {\it Quarterly Journal of Economics},  815-842.}

\item[]{\small \textsc{Eckles, D., B. Karrer, and J. Ugander} (2014),
``Design and Analysis of Experiments in Networks: Reducing Bias from Interference'', unpublished working paper.}

\item[]{\small \textsc{Eicker, F.}, (1967), ``Limit Theorems for Regression
with Unequal and Dependent Errors,'' \textit{Proceedings of the Fifth
Berkeley Symposium on Mathematical Statistics and Probability}, Vol. 1,
59-82, University of California Press, Berkeley.}

\item[]{\small\textsc{Firpo, S.} (2007),
``Efficient Semiparametric Estimation of Quantile Treatment Effects,''{\it Econometrica}, Vol. 75(1): 259-276. 
}

\item {\small \textsc{Fisher, R. A.}, (1925), {\it Statistical Methods for Research Workers}, 1st ed, Oliver and Boyd, London.  }

\item {\small \textsc{Fisher, R. A.}, (1935), {\it Design of Experiments}, Oliver and Boyd.  }

\item {\small \textsc{Fisher, L. et al.}, (1990), ``Intention-to-Treat in Clinical Trials,''  In K.E. Peace (ed.), {\it Statistical Issues in Drug Research and Development,}  New York: Marcel Dekker. }

\item {\small \textsc{Fithian, W., Sun, C., and J. Taylor}, (2015), ``Optimal Inference After
Model Selection,'' \\ http://arxiv.org/abs/1410.2597.}

\item {\small \textsc{Foster, J., and Taylor, J. and S. Ruberg,},  ``Subgroup identification from randomized clinical trial data'', {\it Statistics in medicine}, 30 (24), 2867-2880.}

\item[] {\small \textsc{Frangakis, C., and D. Rubin}, (2002), \textquotedblleft Principal Stratification,\textquotedblright {\it Biometrics}, Vol (1): 21-29. }

\item[] {\small \textsc{Freedman, D.}, (2006), \textquotedblleft Statistical Models for Causality: What Leverage do They Provide,\textquotedblright  {\it Evaluation Review} , Vol 30, 691-713. }

\item[] {\small \textsc{Freedman, D.}, (2008), \textquotedblleft On Regression Adjustmens to Experimental Data,\textquotedblright  {\it Advances in Applied Mathematics} , Vol 30(6), 180-193. }

\item[]{\small\textsc{Gail, M., W. Tian, and S. Piantadosi}, (1988),
\textquotedblleft Tests for No Treatment Effect in Randomized Clinical Trials,\textquotedblright {\it Biometrika}, Vol 75(3): 57-64. }

\item[]{\small \textsc{Gail, M., S. Mark, R. Carroll, S. Green, and D. Pee,} (1996),
\textquotedblleft On Design Considerations and Randomization-based Inference for Coomunity Intervention Trials,\textquotedblright\ {\it Statistics in Medicine}, Vol 15, 1069-1092. }

\item[] {\small \textsc{Glennerster, R.}, (2016),
``The Practicalities of Running Randomized Evaluations: Partnerships, Measurement, Ethics, and Transparency'', 
{\it Handbook of Development Economics}, Banerjee and Duflo (eds.), Elseviers, North Holland.}

\item[] {\small \textsc{Glennerster, R., and K. Takavarasha}, (2013),
{\it Running Randomized Evaluations: A Practical Guide}, Princeton University Press.}

\item[] {\small \textsc{Graham, B., G. Imbens, and G. Ridder}, (2014), 
``Complementarity and aggregate implications of assortative matching: a nonparametric analysis,''
{\it  Quantitative Economics}, Vol.  5(1): 29-66.}

\item[] {\small \textsc{Green, D., and H. Kern}, (2012), {\it Detecting Heterogeneous Treatment Effects in Large-Scale Experiments Using Bayesian Additive Regression Trees}, Public opinion quarterly, 76 (3), 491-511.}

\item[]{\small \textsc{H\"ardle, W.} (2002), {\it Applied Nonparametric Regression Analysis},  Cambridge University Press. }

\item[] {\small \textsc{Hill, J.}, (2011), {\it Bayesian nonparametric modeling for causal inference},  Journal of Computational and Graphical Statistics, 20 (1).}

\item[] {\small \textsc{Hinkelmann, K., and O. Kempthorne.}, (2008), {\it Design and Analysis of Experiments}, Vol. 1, Introduction to Experimental Design, Wiley. }

\item[] {\small \textsc{Hinkelmann, K., and O. Kempthorne.}, (2005), {\it Design and Analysis of Experiments}, Vol. 2, Advance Experimental Design, Wiley.}

\item[]{\small \textsc{Hirano, K., G. Imbens, D. Rubin, and A. Zhou}, (2000),
``Estimating the Effect of Flu Shots in a
Randomized Encouragement Design,'' 
{\it Biostatistics}, 1(1): 69-88.}

\item[]{\small \textsc{Hirano, K. and Porter, J.}, (2009), ``Asymptotics for statistical treatment rules,'' {\it Econometrica}, 1683--1701, 77(5).}

\item[]{\small \textsc{Hodges, J.L. and Lehmann, E.}, (1970).  {\it Basic Concepts of Probability and Statistics,} 2nd ed.  San Francisco: Holden-Day.}

\item[]{\small \textsc{Holland, P.}, (1986), \textquotedblleft Statistics and
Causal Inference\textquotedblright (with discussion), {\it Journal of the American
Statistical Association}, 81, 945-970. }

\item[]{\small \textsc{Hotz J., G. Imbens, and J. Mortimer} (2005),
``Predicting the Efficacy of Future Training Programs Using Past
Experiences,'' {\it Journal of Econometrics}, Vol. 125: 241-270.}

\item[]{\small \textsc{Huber, P.}, (1967), ``The Behavior of Maximum
Likelihood Estimates Under Nonstandard Conditions,'' \textit{Proceedings of
the Fifth Berkeley Symposium on Mathematical Statistics and Probability},
Vol. 1, 221-233, University of California Press, Berkeley.}

\item[]{\small \textsc{Hudgens, M., and M. Halloran,} (2008),
``Toward Causal Inference With Interference'' {\it Journal of the American Statistical Association}, 103(482): 832-842.}

\item[]{\small \textsc{Imai, K. and M. Ratkovic}, ``Estimating Treatment Effect Heterogeneity in Randomized Program Evaluation,''  {\it Annals of Applied Statistics},  7(1), (2013), 443-470.}

\item[]{\small\textsc{Imbens, G.}, (2010),
``Better LATE Than Nothing: Some
Comments on Deaton (2009) and
Heckman and Urzua (2009),'' {\it Journal of Economic Literature}, 399-423.}


\item[]{\small \textsc{Imbens, G., and J. Angrist} (1994),
``Identification and Estimation of Local Average Treatment Effects,''
{\it Econometrica}, Vol. 61, No. 2, 467-476.
}

\item[]{\small \textsc{Imbens, G., and M. Kolesar,} (2015),
``Robust Standard Errors in Small Samples: Some Practical Advice ,''
{\it Review of Economics and Statistics}, forthcoming.
}

\item[] {\small \textsc{Imbens, G., and P. Rosenbaum,}, (2005), \textquotedblleft Randomization Inference with an Instrumental Variable\textquotedblright {\it Journal of the Royal Statistical Society}, Series A, vol 168(1), 109-126.}

\item[]{\small\textsc{Imbens, G.,  and D. Rubin}, (1997a),
``Estimating Outcome Distributions for Compliers
in Instrumental Variable Models,'' {\it Review of
Economic Studies}  64(3): 555--574.}

\item[]{\small\textsc{Imbens, G.,  and D. Rubin}, (1997b),
``Bayesian Inference for Causal Effects in
Randomized Experiments with Noncompliance,'' {\it Annals of Statistics}, Vol. 25, No. 1, 305--327. 
}

\item {\small \textsc{Imbens, G., and D. Rubin,} (2015),
\textquotedblright\ \textit{Causal Inference in Statistics, Social, and Biomedical Sciences: An Introduction}%
, Cambridge University Press.}

\item[] {\small \textsc{Kempthorne, O.}, (1952), {\it The Design and Analysis of Experiments}, Robert Krieger Publishing Company, Malabar, Florida.}

\item[] {\small \textsc{Kempthorne, O.}, (1955), ``The Randomization Theory of Experimental Evidence,'' {\it Journal of the American Statistical Association}, Vol. 509271): 946-967.
 }

\item[]{\small \textsc{Lalonde, R.J.,} (1986), ``Evaluating the Econometric
Evaluations of Training Programs with Experimental Data,'' {\it American
Economic Review}, 76, 604-620. }

\item[]  {\small \textsc{Lee, M.-J.}, (2005), {\it Micro-Econometrics for
Policy, Program, and Treatment Effects} Oxford University Press, Oxford. }

\item[]{\small \textsc{Lehman, E.}, (1974),
{\it Nonparametrics: Statistical Methods Based on Ranks}, Holden-Day, San Francisco. }

\item[] {\small \textsc{Lesaffre, E., and S. Senn.}, (2003), \textquotedblleft A Note on Non-Parametric ANCOVA for Covariate Adjustment in Randomized Clinical Trials,\textquotedblright   {\it Statistics in Medicine}, Vol. 22, 3583-3596.
}

\item{\small\textsc{Liang, K., and S. Zeger}, (1986), ``Longitudinal Data Analysis Using Generalized Linear Models,'' 
{\it Biometrika}, 73(1): 13-22.}

\item {\small \textsc{Lin, W.,} (2013), \textquotedblleft Agnostic Notes on
Regression Adjustments for Experimental Data: Reexamining Freedman's
Critique,\textquotedblright\ \textit{The Annals of Applied Statistics}, Vol.
7:(1): 295--318.}

\item[]{\small \textsc{List, J. Shaikh, A. and Y. Xu} (2016),
``Multiple Hypothesis Testing in Experimental Economics,'' NBER Working Paper No. 21875.}

\item[]{\small \textsc{Liu, L., and M. Hudgens} (2013),
``Large Sample Randomization Inference of Causal Effects in the Presence of Interference,'' {\it Journal of the American Statistical Association}, 288-301.}

\item[]{\small\textsc{List, J., A. Shaikh, and Y. Xu,} (2016),
``Multiple Hypothesis Testing in Experimental Economics,''
NBER Working Paper No. 21875.}

\item[] 
{\small\textsc{Lui, K.} (2011), {\it Binary Data Analysis of Randomized Clinical Trials with Noncompliance}, Wiley, Statistics in Practice. }

\item[]{\small \textsc{Lynn, H., and C. McCulloch}, (1992),
\textquotedblleft When Does it Pay to Break the Matches for Analysis of a Matched-pair Design,\textquotedblright\ {\it Biometrics}, Vol 48, 397-409. }

\item[]{\small \textsc{Manning, W., J. Newhouse, N. Duan, E. Keeler and A.
Leibowitz}, (1987),
``Health Insurance and the Demand for Medical Care: Evidence from a Randomized Experiment''
{\it The American Economic Review}, Vol. 77(3): 251-277}

\item[]{\small \textsc{Manski, C.}, (1990), ``Nonparametric Bounds on Treatment
Effects,'' {\it American Economic Review Papers and Proceedings}, 80,
319-323. }

\item[]{\small \textsc{Manski, C.,} (1993),
``Identification of Endogenous Social Effects: The Reflection Problem,'' {\it Review of Economic Studies}, 60(3):531-542.}

\item {\small \textsc{Manski, C.}, (1996), \textquotedblleft Learning about Treatment Effects from Experiments with Random Assignment of Treatments,\textquotedblright {\it The Journal of Human Resources}, 31(4): 709-73. }

\item[]{\small \textsc{Manski, C.} (2003), {\it Partial Identification of
Probability Distributions}, New York: Springer-Verlag. }

\item[]{\small \textsc{Manski, C.} (2003), ``Statistical treatment rules for heterogeneous populations,'' {\it Econometrica},
 1221--1246, 72(4).}

\item[]{\small \textsc{Manski, C.} (2013), {\it 
Public Policy in an Uncertain World}, Harvard University Press, Cambridge. }

\item[] {\small \textsc{Meager, R.}, (2015),
``Understanding the Impact of Microcredit Expansions: A
Bayesian Hierarchical Analysis of 7 Randomised Experiments''
MIT, Department of Economics.}

\item[] {\small \textsc{McNamee, R.}, (2009), \textquotedblleft Intention to Treat, Per Protocol, As Treated and Instrumental Variable Estimators given Non-Compliance and Effect Heterogeneity,\textquotedblright   {\it Statistics in Medicine}, Vol. 28, 2639-2652. }

\item[]{\small\textsc{Miguel, T., and M. Kremer,} (2004),
``Worms: Identifying Impacts on Education and Health in the Presence of Treatment Externalities,'' {\it Econometrica}, Vol. 72(1): 159-217.}

\item[] 
 {\small \textsc{Morgan, K. and D. Rubin}, (2012), ``Rerandomization to improve covariate balance in experiments,'' {\it Annals of Statistics} Vol. 40(2): 1263-1282.}

\item[] 
 {\small \textsc{Morgan, S. and C. Winship}, (2007), {\it %
Counterfactuals and Causal Inference}, Cambridge University Press, Cambridge. 
}

\item[] 
{\small\textsc{Morton, R., and K. Williams,} (2010), {\it Experimental Political Science and the Study of Causality}, Cambridge  University
Press, Cambridge, MA. }

\item[] 
{\small\textsc{Murphy, K., B. Myors, and A. Wollach,} (2014), {\it Statistical Power Analysis}, Routledge. }

\item[] 
 {\small \textsc{Murray, B.}, (2012), {\it Clustering: A Data Recovery Approach}, (second edition), Chapman and Hall. 
}

\item[] {\small \textsc{Neyman, J.}, (1923, 1990), \textquotedblleft On the
Application of Probability Theory to Agricultural Experiments. Essay on
Principles. Section 9,\textquotedblright translated in {\it Statistical Science}, (with
discussion), Vol 5, No 4, 465--480, 1990.}

\item[] {\small \textsc{Neyman, J. with the cooperation of K. Iwaskiewicz and St. Kolodziejczyk}, (1935), \textquotedblleft Statistical problems in agricultural experimentation \textquotedblright (with discussion) Supplement {\it  Journal of the Royal Statistal Society}, Series B, Vol.  2:107-180. }

\item[] {\small \textsc{Olken, B.}, (2015), \textquotedblleft  Promises and Perils of Pre-Analysis Plans,\textquotedblright  {\it  Journal of Economic Perspectives}, Vol  29(3): 61-80. }

\item[] {\small \textsc{Pearl, J.,} (2000, 2009), {\it Causality: Models, Reasoning
and Inference}, Cambridge, Cambridge University Press. }

\item[] {\small \textsc{Robins, P.}, (1985), ``A Comparison of the Labor Supply Findings from the Four Negative Income Tax Experiments'', {\it The Journal of Human Resources}, Vol. 20(4): 567�582.}

\item[] {\small \textsc{Romano, J., A. Shaikh, and M. Wolf,} (2010), ``Hypothesis Testing in Econometrics'', {\it Annual Review of Economics}, Vol. 2: 75-104.}

\item[] {\small \textsc{Romer, Christina D., and David H. Romer}, (2004), 
\textquotedblleft A New Measure of Monetary Shocks: Derivation and Implications,\textquotedblright 
{\it
The American Economic Review}, Vol. 94, No. 4: 1055-1084.}

\item[]{\small \textsc{Rosenbaum, P.}, (2009), {\it Design of Observational Studies}, 
Springer Verlag, New York.}

\item[]{\small \textsc{Rosenbaum, P.}, (1995, 2002), {\it Observational Studies},
Springer Verlag, New York.}

\item {\small \textsc{Rosenbaum, P.,} (2002), \textquotedblleft Covariance
Adjustment in Randomized Experiments and Observational
Studies,\textquotedblright\ \textit{Statistical Science}, Vol. 17:(3):
286--304.}

\item[] {\small \textsc{Rothstein, J., and T. von Wachter}, (2016), \textquotedblleft   Social Experiments in the Labor Market ,\textquotedblright   Handbook of Experimental Economics. }

\item{\small \textsc{Rubin, D.} (1974), "Estimating Causal Effects of Treatments
in Randomized and Non-randomized Studies," \textit{Journal of Educational
Psychology}, 66, 688-701.}

\item[]
{\small\textsc{Rubin, D.} (1975),  \textquotedblleft Bayesian Inference for Causality: The Importance of Randomization,\textquotedblright {\it Proceedings of the Social Statistics Section of the American Statistical Association }, 233-239. }

\item[]{\small\textsc{Rubin, D. B.}, (1978),  \textquotedblleft Bayesian inference for causal
effects: The Role of Randomization,\textquotedblright {\emph
Annals of Statistics}, 6:34--58.}

\item[]
{\small\textsc{Rubin, D.} (2006),   {\it Matched Sampling for Causal Effects},  Cambridge University Press, Cambridge. }

\item[]
{\small\textsc{Rubin, D.} (2007),  \textquotedblleft The Design Versus the Analysis of Observational Studies for Causal Effects: Parallels with The Design of Randomized Trials,\textquotedblright {\it 
Statistics in Medicine}, Vol. 26(1): 20-30. }

\item {\small \textsc{Samii, C., and P. Aronow,} (2012), \textquotedblleft
On equivalencies between design-based and regression-based variance
estimators for randomized experiments\textquotedblright\ \textit{Statistics
and Probability Letters} Vol. 82: 365--370.}

\item[]{\small \textsc{Sacerdote, B.,} (2001),
``Peer Effects with Random Assignment: results for Dartmouth Roommates,'' {\it Quarterly Journal of Economics}, 116(2):681-704.}

\item {\small \textsc{Schochet, P.,} (2010), \textquotedblleft Is Regression
Adjustment Supported by the Neyman Model for Causal
Inference?\textquotedblright\ \textit{Journal of Statistical Planning and
Inference}, Vol. 140: 246--259.}

\item[]
{\small\textsc{Senn, S.} (1994),  \textquotedblleft Testing for Baseline Balance in Clinical 
Trials,\textquotedblright {\it Statistics in Medicine},  Vol 13, 1715-1726. }

\item[] 
{\small\textsc{Shadish, W., T. Cook, and D. Campbell,} (2002), {\it Experimental and Quasi-experimental Designs for Generalized Causal Inference}, Houghton Mifflin.} 

\item[]
{\small\textsc{Signovitch, J.},  {\it Identifying informative biological markers
in high-dimensional genomic data and clinical trials,} PhD Thesis, Department of Biostatistics, Harvard University, (2007).}

\item[]
{\small\textsc{Su, X., C. Tsai, H. Wang, D. Nickerson, and B. Li}, {\it Subgroup Analysis via Recursive Partitioning},  Journal of Machine Learning Research, 10, (2009), 141-158. }

\item[]
{\small\textsc{Taddy, M., M. Gardner, L. Chen, and D. Draper,}, {\it Heterogeneous Treatment Effects in Digital Experimentation}, Unpublished Manuscript, (2015), arXiv:1412.8563.}

\item[]
{\small\textsc{Tian, L., A. Alizadeh, A. Gentles, and R. Tibshirani}, {\it A Simple Method for Estimating Interactions Between a Treatment and a Large Number of Covariates},  Journal of the American Statistical Association, 109(508), (2014) 1517-1532.}

\item[]{\small \textsc{Ugander, J., B. Karrer, L. Backstrom, and J. Kleinberg} (2013),
``Graph Cluster Randomization: Network Exposure to Multiple Universes,''
{\it Proc. of KDD}. ACM.}

\item[]{\small \textsc{Wager, S. and S. Athey}, {\it Estimation and Inference of Heterogeneous Treatment Effects using Random Forests}
arxiv.org:1510.04342
 (2015).}
 
 \item[]{\small \textsc{Wager, S. and G. Walther} (2015), ``Uniform Convergence of Random Forests via Adaptive Concentration,''
arXiv:1503.06388.}

\item[]{\small \textsc{H. Weisburg, H. and V. Pontes}, {\it Post hoc subgroups in Clinical Trials: Anathema or Analytics?}  Clinical Trials, June, 2015.}

\item[]{\small \textsc{White, H.} (1980), ``A Heteroskedasticity-Consistent
Covariance Matrix Estimator and a Direct Test for Heteroskedasticity,''
\textit{Econometrica}, 48, 817-838.}

\item[] 
{\small\textsc{Wu, J., and Hamada, M.,} (2009), {\it Experiments, Planning, Analysis and Optimization}, Wiley Series in Probability and Statistics.}

\item[]{\small \textsc{Young, A.}, (2016)
``Channelling Fisher: Randomization Tests and the Statistical
Insignificance of Seemingly Significant Experimental Results,''
London School of Economics.}

\item[]{\small \textsc{Zeileis, A., T. Hothorn, and K. Hornik},
{\it Model-based recursive partitioning.}   Journal of Computational and Graphical Statistics, 17(2),  (2008), 492-514.}

\item[]{\small \textsc{Zelen, M.}, (1979),
"A New Design for Randomized Clinical Trials",
{\it New England Journal of Medicine}, 300, 1242--1245. }

\item[]{\small \textsc{Zelen, M.}, (1990),
"Randomized Consent Designs for Clinical Trials: An Update",
{\it Statistics in Medicine}, Vol. 9, 645--656. }

\end{description}

\end{document}